\documentclass[journal]{IEEEtran}

\ifCLASSINFOpdf

\else

\fi

\usepackage{mathrsfs}
\usepackage{amsmath}
\usepackage{amsfonts}
\usepackage{amsthm}
\usepackage{amssymb}
\usepackage{graphicx}
\usepackage{subfigure}
\usepackage{indentfirst}
\usepackage{array}
\usepackage{epstopdf}
\usepackage{cite}
\usepackage{mathrsfs}
\usepackage{enumerate}
\usepackage{bm}
\usepackage[linesnumbered,ruled,vlined]{algorithm2e}
\usepackage{setspace}
\usepackage{multirow}
\usepackage{verbatim}
\usepackage{stfloats}
\usepackage{makecell}
\usepackage{cancel}
\usepackage{float}
\usepackage{threeparttable}
\usepackage[dvipsnames]{xcolor}
\usepackage[colorlinks=true, linkcolor=blue, citecolor=blue]{hyperref}

\definecolor{lightgray}{rgb}{0.83, 0.83, 0.83}

\definecolor{revised}{HTML}{b71a3b}

\SetKwComment{Comment}{//}{}
\SetKwBlock{Begin}{Function}{end}

\begin{document}
	
	\title{Rate-Distortion-Perception Controllable Joint Source-Channel Coding for High-Fidelity Generative Communications}
	
	\author{Kailin~Tan,~\IEEEmembership{Graduate Student Member,~IEEE},
		Jincheng~Dai,~\IEEEmembership{Member,~IEEE},
		Zhenyu~Liu,~\IEEEmembership{Member,~IEEE},
		Sixian~Wang,~\IEEEmembership{Member,~IEEE},
		Xiaoqi~Qin,~\IEEEmembership{Senior Member,~IEEE},
		Wenjun~Xu,~\IEEEmembership{Senior Member,~IEEE},
		Kai~Niu,~\IEEEmembership{Member,~IEEE},
		Ping~Zhang,~\IEEEmembership{Fellow,~IEEE}

		\thanks{This work was supported in part by the National Natural Science Foundation of China under Grant 62293481, Grant 62371063, Grant 62201089, Grant 92267301, in part by the Natural Science Foundation of Beijing Municipality under Grant 4222012, and in part by the Program for Youth Innovative Research Team of BUPT No. 2023YQTD02, in part by BUPT Excellent Ph.D. Students Foundation CX2023118.}
		
		\thanks{Kailin Tan, Jincheng Dai, Sixian Wang, and Kai Niu are with the Key Laboratory of Universal Wireless Communications, Ministry of Education, Beijing University of Posts and Telecommunications, Beijing 100876, China.}
		
		\thanks{Zhenyu Liu is with the 5GIC and 6GIC, Institute for Communication Systems, University of Surrey, GU2 7XH Guildford, United Kingdom.}
		
		\thanks{Xiaoqi Qin, Wenjun Xu, and Ping Zhang are with the State Key Laboratory of Networking and Switching Technology, Beijing University of Posts and Telecommunications, Beijing 100876, China.}
		
		\vspace{-1em}
	}
	
	\maketitle

	\begin{abstract}
		End-to-end image transmission has recently become a crucial trend in intelligent wireless communications, driven by the increasing demand for high bandwidth efficiency. However, existing methods primarily optimize the trade-off between bandwidth cost and objective distortion, often failing to deliver visually pleasing results aligned with human perception. In this paper, we propose a novel rate-distortion-perception (RDP) jointly optimized joint source-channel coding (JSCC) framework to enhance perception quality in human communications. Our RDP-JSCC framework integrates a flexible plug-in conditional Generative Adversarial Networks (GANs) to provide detailed and realistic image reconstructions at the receiver, overcoming the limitations of traditional rate-distortion optimized solutions that typically produce blurry or poorly textured images.
		Based on this framework, we introduce a distortion-perception controllable transmission (DPCT) model, which addresses the variation in the perception-distortion trade-off. DPCT uses a lightweight spatial realism embedding module (SREM) to condition the generator on a realism map, enabling the customization of appearance realism for each image region at the receiver from a single transmission.
		Furthermore, for scenarios with scarce bandwidth, we propose an interest-oriented content-controllable transmission (CCT) model. CCT prioritizes the transmission of regions that attract user attention and generates other regions from an instance label map, ensuring both content consistency and appearance realism for all regions while proportionally reducing channel bandwidth costs.
		Comprehensive experiments demonstrate the superiority of our RDP-optimized image transmission framework over state-of-the-art engineered image transmission systems and advanced perceptual methods.

	\end{abstract}
	
	
	\IEEEpeerreviewmaketitle
	
	\section{Introduction}\label{section_introduction}
	
	The rapid expansion of ultra-large-scale image/video transmission applications in camera phones and extended reality devices continues to drive the demand for efficient transmission of large media data under limited bandwidth conditions. Traditional communication systems, based on the source-channel separation paradigm, utilize rate-distortion theory for source coding and channel coding theory for transmission. These systems aim to minimize the size of source data under a distortion constraint while ensuring reliable data transmission over noisy channels. However, they can cause significant bandwidth waste due to their focus on global bit information rather than critical semantic information. To address this inefficiency, recent advancements in deep learning have inspired data-driven solutions that extract semantic feature information \cite{9955525, dai2021semantic, 10026795} and implement joint source-channel coding (JSCC) \cite{DJSCC, DJSCCF, DJSCCL, saidutta2021joint, dai2022nonlinear, dai2022adaptive} for end-to-end communications.
	
	Initial works in DeepJSCC \cite{DJSCC, DJSCCF, DJSCCL} typically employ an autoencoder architecture that directly maps source information to channel-input symbols, considering wireless channel impairments to minimize end-to-end distortion metrics (e.g., the mean-squared error (MSE)). It has been demonstrated that DeepJSCC can outperform classical separation schemes, such as JPEG combined with LDPC channel coding \cite{3gpp} for small images \cite{DJSCC}. In \cite{yang2023witt, yang2024swinjscc}, the authors proposed SwinJSCC, which is built on the Swin Transformer architecture, and showed that SwinJSCC outperforms BPG \cite{BPG} (compatible with HEVC intra coding) combined with 5G LDPC channel coding for medium-sized images. However, these works only optimize a single distortion metric with a fixed rate.
	
	To enhance transmission efficiency in variable-rate coding (VRC) under varying channel conditions and available bandwidth, the nonlinear transform source-channel coding (NTSCC) approach \cite{saidutta2021joint, dai2022nonlinear, dai2022adaptive, wang2023improved} integrates a learned entropy model in the latent space to optimize the trade-off between transmission rate and distortion metrics end-to-end. This rate-distortion (RD) optimized JSCC framework has demonstrated significant coding gains over distortion-optimized JSCC approaches, exhibiting comparable or superior end-to-end RD performance \cite{wang2023improved} compared to state-of-the-art (SOTA) engineered source and channel codecs, such as VTM \cite{bross2021developments} combined with 5G LDPC. However, the reconstructed images from these methods, despite reduced distortion, which can degrade human perceptual quality due to their pixel-wise averages of plausible solutions. As the transmission rate decreases, the first aspect to degrade in the reconstructed images is the texture details, resulting in an increasingly smoother appearance, and eventually leading to damage to the image contents. However, high perceptual quality indicates high-fidelity reconstructions that possess both ``appearance realism'' and ``content consistency'', where ``appearance realism'' requires sharp and appropriate textures to provide a visually pleasing and natural appearance.
	
	To improve perceptual quality while ensuring limited distortion in the wireless transmission of images, it is essential to clarify the differences and connections between distortion and perception. Specifically, ``distortion'' refers to the dissimilarity between the reconstructed signal and the source signal, while ``perceptual quality'' is formalized as a divergence between the distribution of source signals and the distribution of reconstructed signals \cite{blau2018perception}. Furthermore, Blau and Michaeli \cite{blau2019rethinking} have theoretically revealed the trade-off between perception and distortion: for a fixed rate, reducing distortion alone can lead to poor perceptual quality, whereas accepting some distortion can enhance perceptual quality. Leveraging this principle, we propose to upgrade the optimization goal of the JSCC framework to incorporate the triple trade-off of ``rate-distortion-perception (RDP)'' to meet the demand for high perceptual quality in end-to-end communication systems.
	
	When we follow the RDP theory to develop the generative JSCC framework, we usually face different practical deployment challenges due to different available bandwidth:
	\begin{itemize}
		
		\item \textbf{Personalized transmission for various perception and distortion demands in affordable bandwidth}: Given feasible bandwidth limitations, different users may have varying perception and distortion requirements for the same visual data source; for example, individuals may have personalized areas of interest in the same image or video stream. To meet these variable demands, the RDP-optimized framework must be distortion-perception controllable. This means the model should be capable of decoding either a realistic reconstruction (fine-grained details for human vision), a smooth one (minimal details for distortion demands), or anything in between. While using multiple models for each distortion and perception trade-off can meet this requirement, it incurs significant training and deployment costs. Therefore, it is desirable to utilize a single model capable of handling these variations from a single received signal in one-shot transmission.
		
		\item \textbf{Interest-oriented generative transmission in scarce bandwidth}: When bandwidth is extremely limited, such as in emergency communication, it becomes challenging to ensure content consistency while reproducing realistic details using extremely low rates. Note that, not all information within the media data is needed or equally important for different users; for example, in emergency rescue, medical personnel might focus more on injuries, while telecommunication maintenance staff might prioritize base station damage. This recognition motivates the development of interest-oriented implementations, where communication success means conveying personalized critical information to meet the receiver’s interest, while lower-priority information can rely on generative models to produce realistic outcomes. 
	\end{itemize}

	\begin{figure}[t]
		\setlength{\abovecaptionskip}{0cm}
		\setlength{\belowcaptionskip}{0cm}
		\centering{\includegraphics[scale=0.227]{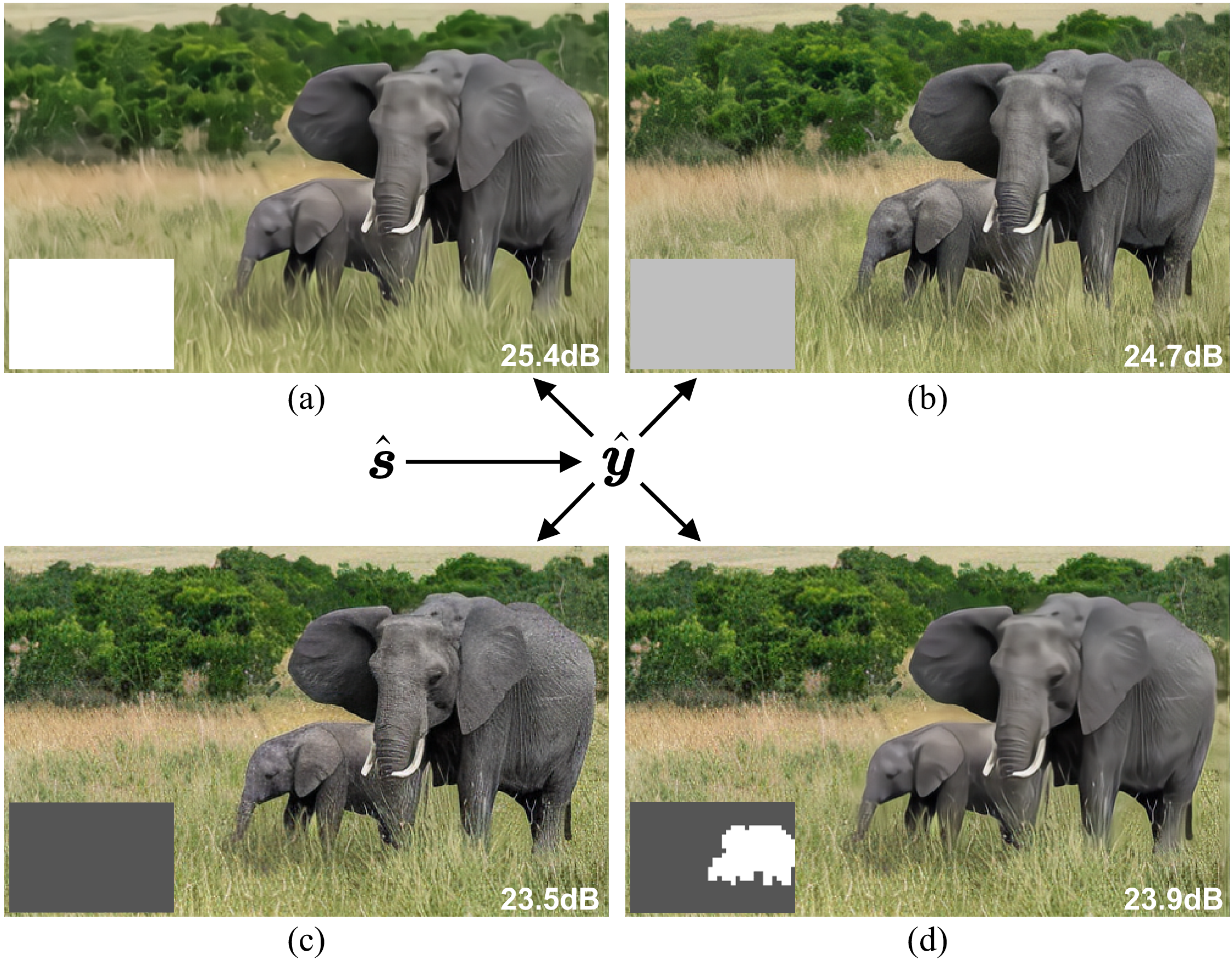}}
		\caption{Flexible DP control using our DPCT model: decoding multiple reconstructions with different distortion-perception trade-offs from a single received signal $\bm{\hat{s}}$ in one-shot transmission. The realism map $\bm{\beta}$ (bottom left of each reconstruction) enables customization of the amount of details for per image region, with darker colors indicating more details. Fig. \ref{fig1} (a), (b), and (c) show the change of preference from distortion to perception by global control mode, and Fig. \ref{fig1} (d) demonstrates the spatial control mode (low distortion for the elephant and high perception for other regions).}
		\label{fig1}
		\vspace{0em}
	\end{figure}
	
	Consequently, in this paper, we develop a novel \emph{RDP} jointly optimized JSCC framework to facilitate perception-oriented end-to-end communications. Our proposed RDP-JSCC framework incorporates a flexible plug-in conditional Generative Adversarial Networks (GANs) \cite{goodfellow2014generative} that can be efficiently integrated with existing RD-based models to enhance human perception. Based on the developed framework, we propose two RDP-JSCC implementations: a distortion-perception controllable transmission (DPCT) model for bandwidth-sufficient scenarios and an interest-oriented content-controllable transmission (CCT) model for bandwidth-scarce scenarios. We demonstrate the superiority of our RDP-optimized solutions in improving perceptual quality, and a visual demo of DPCT is presented in Fig. \ref{fig1}.
	
	Our main contributions are summarized below.
	
	\begin{itemize}
		\item To improve perceptual quality while ensuring limited distortion in wireless image transmission, we develop a novel generative JSCC framework (RDP-JSCC) that jointly optimizes transmission rate, image distortion, and human perception. The proposed RDP-JSCC framework is compatible with existing RD-JSCC frameworks by integrating a flexible plug-in conditional GAN for perception enhancement and a triple RDP optimization objective that includes transmission rate, distortion loss, and GAN loss. To the best of our knowledge, this is the first JSCC framework to achieve RDP joint optimization.
		
		\item Based on the proposed framework, we introduce a DPCT model to manage the personalized distortion-perception trade-off in bandwidth-affordable scenarios. To efficiently enhance the DP trade-off, our DPCT model conditions the generator on a \emph{realism map} using a lightweight \emph{spatial realism embedding module (SREM)}. Our approach enables customization of the appearance realism for each image region only at the receiver from a signal received signal. Experimental results show that our DPCT model dominates existing transmission methods in terms of perceptual performance, while its distortion performance is only slightly inferior to or even close to VTM combined with 5G LDPC.
		
		\item Unlike existing works that primarily encode all features with comparable priority, leading to content inconsistency and loss of important details in bandwidth-scarce scenarios, we propose an interest-oriented CCT model that can controllably generate any concerned content completely. By fully leveraging the biased property \cite{cai2019end} of user perception, we prioritize the transmission of interested regions that attract user attention and generate other regions from an instance label map, ensuring content consistency and appearance realism for all regions. Compared to the coding overhead of the image, the overhead caused by the instance label map is minimal. More importantly, the size of transmitted symbols is reduced in proportion to the regions generated from the instance label map, ultimately resulting in substantial channel bandwidth savings.
	\end{itemize}
	
	The rest of this paper unfolds as follows. Section \ref{RDP_JSCC} introduces the proposed RDP-JSCC framework. Then, Section \ref{DPCT} and Section \ref{CCT} present the implementation details of DPCT and CCT, respectively. Experimental results, presented in Section \ref{Experimental_Results}, elucidate the performance improvement achieved by DPCT and CCT. We conclude this paper in Section \ref{Conclusion}.
	
	\section{The RDP-JSCC Framework}\label{RDP_JSCC}
	\begin{figure}[t]
		
		\setlength{\abovecaptionskip}{0cm}
		\setlength{\belowcaptionskip}{0cm}
		\centering{\includegraphics[scale=0.43]{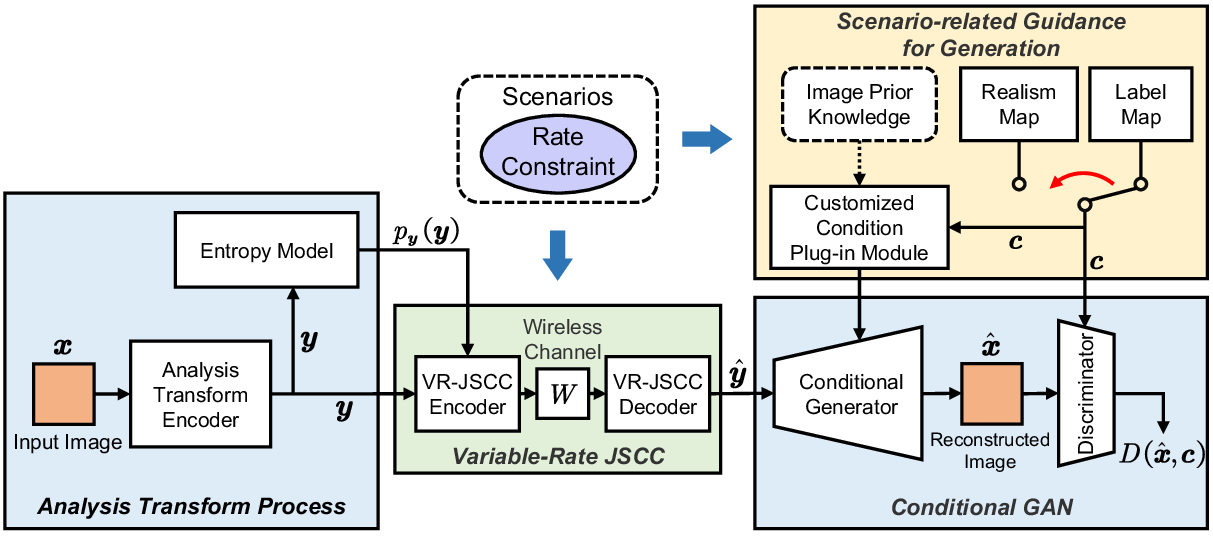}}
		\caption{Architecture of our RDP-JSCC framework, which includes analysis transform process (an analysis transform encoder and an entropy model), variable-rate JSCC, a conditional GAN, and a customized condition plug-in module. $\bm{c}$ is the conditional information, and it varies with different rate scenarios. The discriminator $D$ is only operated during the training phase.}
		\label{fig2}
		\vspace{0em}
	\end{figure}
	In this section, we propose a novel rate-distortion-perception jointly optimized JSCC framework, which includes analysis transform process, variable-rate JSCC, a conditional GAN, and a customized condition plug-in module as shown in Fig. \ref{fig2}.
	
	\subsection{Variable-Rate JSCC Transmitter: Towards High Transmission Efficiency}\label{Tran_efficiency}
	To achieve high transmission efficiency, our framework includes three modularized components: a pair of nonlinear transforms, an entropy model, and variable-rate JSCC (VR-JSCC). Specifically, given a source image $\bm{x}\in \mathbb{R}^{3 \times W \times H}$ ($W$ and $H$ are the width and height of $\bm{x}$), the analysis transform encoder $g_a$ and synthesis transform decoder $g_s$ is responsible for transforming $\bm{x}$ into latent features $\bm{y} \in \mathbb{R}^{c \times w \times h}$ ($w$, $h$, and $c$ are the width, height, and channel numbers of $\bm{y}$) and recovering the reconstructed latent features $\bm{\hat{y}}$ back into the source image, respectively. Note that $g_s$ also functions as a generator in Fig. \ref{fig2}.
	
	In the latent space, the learned entropy model $\varepsilon$ estimates the parameterized distributions of $\bm{y}$, i.e., $p_{\bm{y}}\left( \bm{y} \right)$. The transmitter leverages $p_{\bm{y}}\left( \bm{y} \right)$ to calculate the entropy of a quantized version of $\bm{y}$, which is utilized to estimate the bandwidth required for transmitting $\bm{y}$. Concurrently, $\boldsymbol{y}$ is encoded by the VR-JSCC encoder $f_e$ into the sequence of channel symbols $\boldsymbol{s} \in \mathbb{R}^k$, and the receiver gets a sequence ${\boldsymbol{\hat s}} = W( \boldsymbol{s} )$. In this paper, we consider the general fading channel model, such that this process can be represented as ${\boldsymbol{\hat s}} = W( \boldsymbol{s} ) = \bm{h} \odot \boldsymbol{s} + \boldsymbol{n}$, where $\odot$ denotes the element-wise product, $\bm{h}$ is the channel gain vector, and $\boldsymbol{n}$ is the noise vector. The VR-JSCC decoder $f_d$ recovers the reconstructed latent features $\bm{\hat{y}}$ from $\bm{\hat{s}}$. In total, the forward procedure of transmission is outlined as:
	\begin{equation}\label{eq_trans_process}
		\begin{aligned}
			& {{\boldsymbol{x}}} \xrightarrow{{{g_a}( \cdot)}} {{\boldsymbol{y}}}
			\xrightarrow{{{f_e}( \cdot)}} {{\boldsymbol{s}}}
			\xrightarrow{{{W}( \cdot )}} {{\boldsymbol{\hat s}}}
			\xrightarrow{{{f_d}( \cdot)}} {{\boldsymbol{\hat y}}}
			\xrightarrow{{{g_s}( \cdot)}} {{\boldsymbol{\hat x}}} \\
			& \text{with the entropy model~} \boldsymbol{y}\xrightarrow{\varepsilon \left( \cdot \right)}\boldsymbol{\varPhi }\xrightarrow{}p_{\boldsymbol{y}}\left( \boldsymbol{y} \right),
		\end{aligned}
	\end{equation}
	where $\boldsymbol{\varPhi }$ is the distribution parameters of $\bm{y}$. The system efficiency is quantified by the channel bandwidth ratio (CBR) $\rho =k/(3 \times W \times H)$.
	
	The optimization objective for the RD trade-off without perception can be expressed as:
	\begin{equation}\label{loss_RD_func}
		\begin{aligned}
			& \mathcal{L}_{RD} = \mathbb{E}_{\boldsymbol{x}\sim p_{\boldsymbol{x}}} \Big( \lambda \underbrace{\big( -{\eta} \log{p_{\boldsymbol{ y}}(\boldsymbol{ y})} \big)}_{\text{transmission rate}} + \underbrace{d(\boldsymbol{x},\boldsymbol{\hat{x}})}_{\text{distortion metric}}\Big),
		\end{aligned}
	\end{equation}
	where the weight factor $\lambda$ determines the trade-off between the transmission rate and the distortion metric. The scaling factor $\eta>0$ correlates the estimated entropy to the channel bandwidth cost. Specifically, each embedding $\bm{y}_i$ is a $c$-dimensional vector, and the bandwidth cost $k_{i}$ for transmitting $\bm{y}_i$ is
	\begin{equation}\label{eq_channel_bandwidth_cost_cal}
		{k}_{i} = Q\Big( -{\eta} \log{p_{\bm{y}_i}(\bm{y}_i)} \Big),
	\end{equation}
	where $Q$ represents the scalar quantization function involving $2^{q}$ ($q = 1,2,\dots$) integers with the quantization value set $\mathcal{V} = \{ v_1, v_2, \dots, v_{2^{q}} \}$. Hence, the receiver can know the bandwidth allocated to every $\bm{y}_i$ by transmitting side information with the predefined $q$ bits. With the VR-JSCC, every $\bm{y}_i$ is adaptively mapped to a ${k}_{i}$-dimensional channel symbol vector $\bm{s}_i$ guided by a rate allocation vecotr $\bm{k}=\left[ k_1,\cdots ,k_l \right] $, where $l=w\times h$. 
	
	\subsection{Generative Receiver: Aligned with Human Perception}\label{GAN_HP}
	
	To optimize perception together with the transmission rate and the distortion jointly in our framework, we introduce conditional GANs \cite{goodfellow2014generative, mirza2014conditional}, widely used for deep generative models, to learn the distribution of prior knowledge. 
	
	In standard conditional GANs, every data point $\bm{x}$ is linked with additional information $\bm{c}$, while the joint distribution $p_{\bm{x},\bm{c}}$ remains unknown. The training procedure involves two competing networks. A generator $G$ conditioned on $\bm{c}$ transforms latent samples $\bm{z}$ from a predetermined known distribution $p_{\bm{z}}$ into samples from $p_{\bm{x}|\bm{c}}$, while a discriminator $D$ maps the input $(\bm{x}, \bm{c})$ to the probability that it is a sample from $p_{\bm{x}|\bm{c}}$ rather than from the output of $G$. The target is for $G$ to ``deceive'' $D$ into believing its samples are real. By holding $\bm{c}$ constant, the non-saturating loss can be optimized as detailed in \cite{goodfellow2014generative}:
	\begin{equation}\label{eq_G_condGANs}
		\begin{aligned}
			&\mathcal{L} _G=\mathbb{E} _{\bm{z}\sim p_{\bm{z}}}\left[ -\log \left( D\left( G\left( \bm{z},\bm{c} \right) ,\bm{c} \right) \right) \right], \\
		\end{aligned}
	\end{equation}
	\begin{equation}\label{eq_D_condGANs}
		\begin{aligned}
			&\mathcal{L} _{D}=\mathbb{E} _{\bm{z}\sim p_{\bm{u}}}\left[ -\log \left( 1-D\left( G\left( \bm{z},\bm{c} \right) ,\bm{c} \right) \right) \right] \\
			&\,\,\,\,\,\,\,\,\,\,\,\,\,\,\,\,+\mathbb{E} _{\bm{x}\sim p_{\bm{x}|\bm{c}}}\left[ -\log \left( D\left( \bm{x},\bm{c} \right) \right) \right]. \,\,
		\end{aligned}
	\end{equation}
	
	At the generative receiver of our framework, we aim to generate high perceptual quality images from the latent feature $\bm{\hat{y}}$. However, $\bm{\hat{y}}$ contains the information of the real image $\bm{x}$ and is affected by channel interference, while transmission rate and distortion are also crucial factors to consider. Obviously, the triple dynamic trade-off under wireless channel clearly goes beyond the equation \eqref{eq_G_condGANs} in standard conditional GANs. Therefore, we must design new compatible structures and new optimization objectives to achieve joint optimization. Specifically, we upgrade the synthesis transform decoder $g_s$ into a generator to reconstruct the image $\bm{\hat{x}}$. Crucially, we also provide conditional information $\bm{c}$ for $g_s$ and $D$ to guide model generation capability by conditioning the distribution of $\bm{x}$ on $\bm{c}$ (i.e., $p_{\boldsymbol{x}|\boldsymbol{c}}$). A discriminator $D$ assesses how likely $\bm{\hat{x}}$ is sampled from the true distribution $p_{\bm{x}|\bm{c}}$. Under such a framework, the goal of $g_s$ includes two: reconstructing the transmitted image $\bm{x}$ to a specific extent; generating $\bm{\hat{x}}$ that aligns with the real conditional distribution $p_{\bm{x}|\bm{c}}$. Therefore, by merging equations \eqref{eq_G_condGANs}, \eqref{eq_D_condGANs} and \eqref{loss_RD_func}, the objective for our RDP-JSCC framework is expressed as
	\begin{equation}\label{loss_F}
		\begin{aligned}
			&\mathcal{L} _{\mathcal{F}}=\mathbb{E} _{\boldsymbol{x}\sim p_{\boldsymbol{x}}}\Big( \underbrace{ d\left( \boldsymbol{x},\boldsymbol{\hat{x}} \right)}_{\text{distortion metric}} +\lambda \underbrace{ \left( -\eta \log p_{\boldsymbol{y}}\left( \boldsymbol{y} \right) \right)}_{\text{transmission rate}}+ \\ &\,\,\,\,\,\,\,\,\,\,\,\,\,\,\,\,  \beta \underbrace{ \left( -\log D\left( \boldsymbol{\hat{x}},\boldsymbol{c} \right)	\right)}_{\text{perception metric}} \Big),
		\end{aligned}
	\end{equation}
	\begin{equation}\label{loss_D}
		\begin{aligned}
			&&\mathcal{L} _{D}^{\prime}=\mathbb{E} _{\boldsymbol{x}\sim p_{\boldsymbol{x}}}\left( -\log \left( 1-D\left( \boldsymbol{\hat{x}},\boldsymbol{c} \right) \right) -\log D\left( \boldsymbol{x},\boldsymbol{c} \right) \right),
		\end{aligned}
	\end{equation}
	where $\bm{\hat{x}}=g_s \left( \bm{\hat{y}}, \bm{c} \right)$, and both $\lambda>0$ and $\beta>0$ act as balance factors: the former between the transmission rate $R$ and the combined distortion metric with perception metric, and the latter between distortion metric itself and perception metric. The full set of neural network modules optimized by \eqref{loss_F} is symbolized by $\mathcal{F}$, which can be detailed as $\mathcal{F} =\left( g_a,g_s, f_e,f_d, \varepsilon \right)$. Training involves a dual-step approach: firstly, $\mathcal{F}$ parameters are adjusted for $\mathcal{L} _{\mathcal{F}}$ minimization, followed by adjusting $D$ parameters to minimize $\mathcal{L} _{D}$.
	
	\subsection{Discussion on the Properties of RDP-JSCC Framework}\label{Insights}
	
	A pivotal factor in the practical implementation of the RDP-JSCC framework in different scenarios is the transmission rate term, defined as $R=-\eta \log p_{\boldsymbol{y}}\left( \boldsymbol{y} \right)$ ($\eta > 0$). 
	
	We first discuss two extreme scenarios:
	\begin{itemize}
		\item Allowing $R$ to be large enough ($k_i\rightarrow \infty $ and $\lambda=0$), makes the practical channel bandwidth to be infinite. In this case, $\mathcal{F}$ is capable of reconstructing $\bm{x}$ nearly losslessly over the noise channel, such that the distortion term would disappear. Consequently, the divergence between $p \left( \bm{x} \right)$ and $p \left( \bm{\hat{x}} \right)$ would also disappear, rendering the GAN loss moot.
		
		\item Conversely, if $R\rightarrow 0$ (using $\lambda = \infty $, $k_i\rightarrow 0$), $\bm{\hat{y}}$ would lost all semantics about the real image $\bm{x}$. In this setting, $\bm{\hat{y}}$ could be random and independent of $\bm{x}$, and the objective becomes akin to a standard GAN combined with a distortion term, serving as a restraint.
	\end{itemize}
	
	From these two scenarios, we can find that the transmission rate $R$ closely affects the role of GANs in image reconstruction. In our framework, we always impose restrictions on $R$, thereby ensuring that the GAN loss neither becomes moot nor serves as a standard GAN. Thus, it needs to balance the distortion term and the GAN loss.
	
	Most previous works constrain $R$ to an affordable range, allowing for almost lossless transmission of contents but without preserving texture details. In this scenario, the distortion term ensures ``content consistency'', while the GAN only needs to generate plausible textures to ensure ``appearance realism''. Besides, $\beta$, as a ``realism factor'', determines the amount of texture generated. Intuitively, when $\beta=0$, the optimization objective \eqref{loss_F} simplifies to \eqref{loss_RD_func}. For a fixed $R$, increasing $\beta$ shifts the model's preference towards perception at the cost of higher distortion. In the field of neural image compression (NIC), \cite{zhang2021universal} implemented a compression scheme with one encoder and multiple decoders (each for a different $\beta$), making it approximately universal across the DP trade-off, thereby reducing the need to design a new encoder for each DP objective. The work \cite{agustsson2023multi} extends this idea to high-resolution image compression. These inspire us to develop a more universal transmission model using a single decoder effective for any $\beta$ to meet various perception and distortion demands in personalized transmission, which we explore in Section \ref{DPCT}.
	
	However, when $R$ cannot carry all contents, especially when some contents are completely lost, how does our framework cater to human perception? In the field of JSCC, this remains an unexplored issue. For instance, the sender does not transmit the channel symbols corresponding to trees to meet scarce bandwidth, and the information related to trees is completely lost. To address this issue, we study in Section \ref{CCT} how to boost GAN to generate contents consistent with the source while producing realistic details. In this way, detailed ``green trees'' can be fully generated, with slightly different shapes, but visually imperceptible.
	
	Consequently, to address the variation in the prior distribution that the GAN needs to learn when $R$ changes, we provide a conditional plug-in module to guide the GAN in generating the required information more accurately. This module can be customized for different scenarios. In the following sections, we will detail the DPCT model and the CCT model, which are custom-designed implementations of RDP-JSCC for scenarios with affordable bandwidth and scarce bandwidth, respectively.
	
	\begin{figure*}[t]
		\setlength{\abovecaptionskip}{0cm}
		\setlength{\belowcaptionskip}{0cm}
		\centering{\includegraphics[scale=0.49]{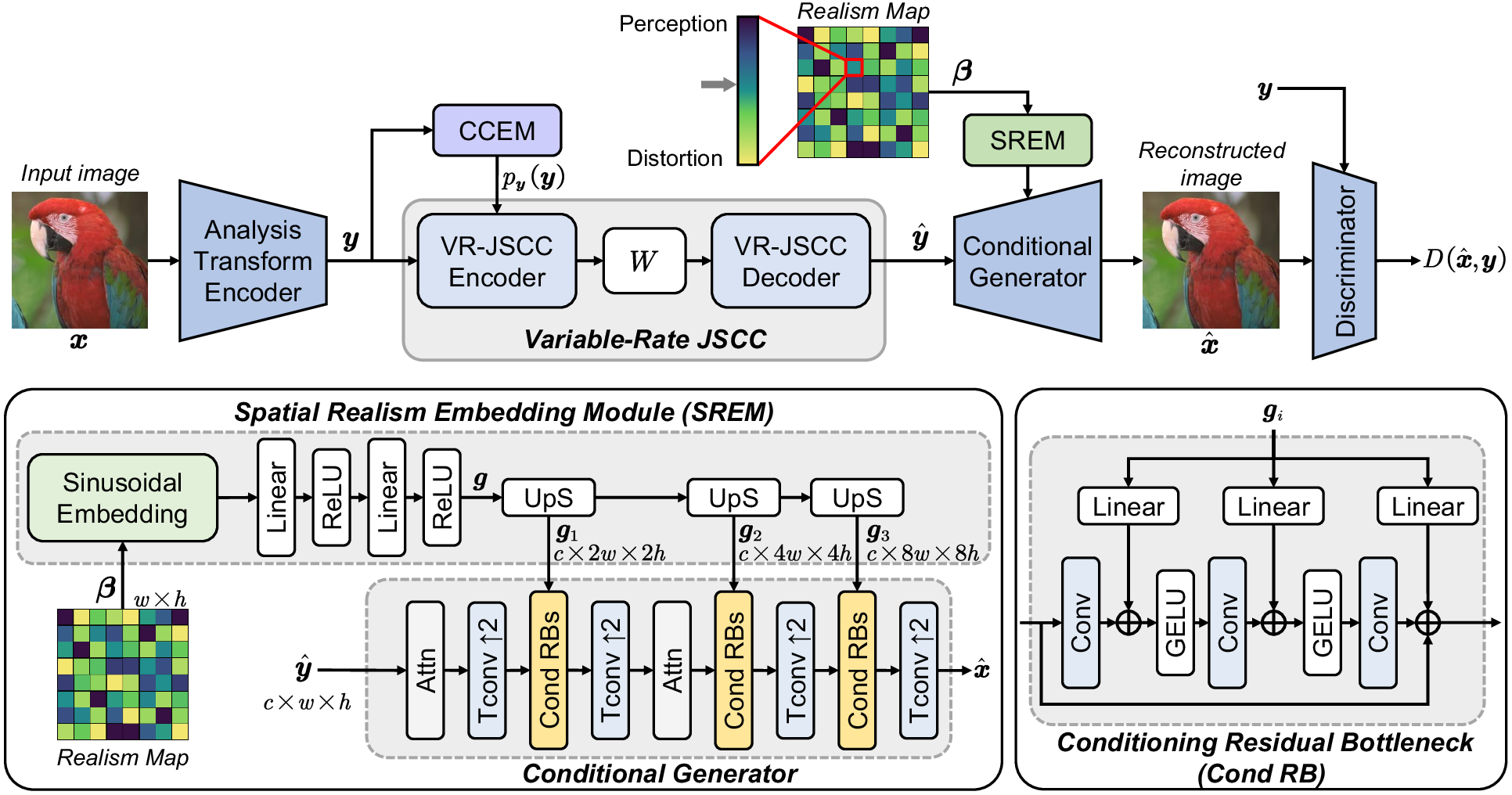}}
		\caption{Network architectures of our DPCT. We design a \emph{spatial realism embedding module (SREM)} to condition $g_s$ on the realism map $\bm{\beta}$. The details of SREM and the conditional generator are given, where \emph{UpS} represents an upsampling operation with a factor of $2$, \emph{Attn} is the attention module, and each \emph{Tconv} is a transposed convolution layer with a stride of 2. \emph{Cond RBs} denotes $3$ stacked \emph{conditioning residual bottleneck (Cond RB)} modules, and the global features $\bm{g}_i$ are fed into every Cond RB.}
		\label{fig3}
		\vspace{0em}
	\end{figure*}
	
	\section{DPCT: Flexible Control for DP Trade-Off under Affordable Bandwidth}\label{DPCT}
	
	Although affordable bandwidth can prevent over-distortion in reconstructed images, users in lossy image transmission can have varying demands—some prioritizing human perception and others requiring low distortion. To simultaneously meet these diverse demands, we propose the DPCT model, which flexibly controls the DP trade-off with one-shot transmission. This control can be applied uniformly across the entire image or even vary spatially with the help of control information from the realism map.
	
	\subsection{Realism Map-based Control on the DP Trade-off}\label{realism_map_conditioning}
	We set a \emph{realism map} $\bm{\beta} \in \mathbb{R}^{w \times h}$ that has the same spatial dimension with $\bm{y}$. Each element $\beta _{i,j}\in \left[ 0,\beta _{\max} \right] $ of $\bm{\beta}$ specifies where and how many details should be generated. Our proposed DPCT can support two typical control modes during the inference phase:
	\begin{itemize}
		\item \textbf{Global control mode}: Setting $\bm{\beta }=b\cdot \mathbf{J}_{w\times h}$, where $b\in \left[ 0,\beta _{\max} \right]$ and $\mathbf{J}_{w\times h}$ represents an $w\times h$ matrix with all elements equal to $1$. If $b$ is a precise value, for example, $b=2$, we simplify this representation to $\bm{\beta}=2$. This represents global control over the entire image, which PDJSCC \cite{wang2022perceptual} can only achieve by training multiple models with different $\beta$ and multiple transmissions.
		
		\item \textbf{Spatial control mode}: Setting $\bm{\beta}$ spatially non-uniformly enables customization of the amount of details per image region. With an instance label map, we can specify the locations of each instance, such as trees, sky, text, etc. Thus, the receiver can subjectively adjust for each instance.
	\end{itemize}
	
	Figure \ref{fig3} illustrates the architecture of our DPCT model, and our goal is to train this model to work well for any $\bm{\beta}$ under one-shot transmission, without significantly increasing the complexity compared to the procedure \eqref{eq_trans_process}. To realize this, on the one hand, we only condition the generator $g_s$ on the realism map $\bm{\beta}$, thereby one received signal $\bm{\hat{s}}$ can be used for decoding any RD trade-off. On the other hand, we leverage $\bm{\beta}$ to spatially weight the perception term in the optimization objective \eqref{loss_F}, guiding the model to optimize specific RD trade-offs across different spatial regions. Standard conditional GANs are refined to fit this model, $G$ and $D$ in \eqref{eq_G_condGANs} and \eqref{eq_D_condGANs} are formulated as
	\begin{equation}\label{eq_G_D_refine_DPCT}
		\hspace{-0.4in}
		\begin{aligned}
			G\leftarrow g_s\left( \bm{\hat{y}}, \bm{\beta} \right) ,\,\,\,\, D\leftarrow D\left( g_s\left( \bm{\hat{y}}, \bm{\beta} \right) ,\boldsymbol{y} \right),
		\end{aligned}
	\end{equation}
	where we take $\bm{\beta}$ and $\bm{y}$ as the conditional information for $D$ and $g_s$ respectively.
	
	To condition $g_s$ on $\bm{\beta}$, we propose a \emph{spatial realism embedding module (SREM)} to embed the control information of $\bm{\beta}$ into $g_s$, as demonstrated in Fig. \ref{fig3}. Besides, for the analysis transform encoder $g_a$ and conditional generator $g_s$, we draw on the ``ELIC'' architectures, which is widely used for NIC works \cite{he2022elic}. We widen its bottleneck dimension to $256$ to enhance the model's expressive capacity. Our entropy model $\varepsilon $ follows the design of the checkerboard context entropy model (CCEM) in \emph{NTSCC+} \cite{wang2023improved}, with details available in \cite{wang2023improved}. Our VR-JSCC also employs the architectures based on swin transformer \cite{wang2023improved}. Besides, our discriminator $D$ adopts a network structure akin to that in PatchGAN \cite{li2016precomputed}. For simplicity, the structures of some modules in Fig. \ref{fig3} are simplified, but this does not affect our discussion.
	
	\subsection{Spatial Realism Embedding Module}\label{SREM}
	The goal of SREM is to embed a continuous-value map into multi-scale feature spaces of the generator. As shown in Fig. \ref{fig3}, we first derive global features $\bm{g}$ by calculating sinusoidal embedding features and connecting them with a two-layer MLP, and this is inspired by Transformer position embedding \cite{vaswani2017attention}. Particularly, given a maximum sinusoidal period $p_{\max}$, we can obtain a constant frequency vector $\bm{\nu}^{freq} \in \mathbb{R}^{c}$ and its every element $\nu^\mathrm{freq}_i$ can be calculated by
	\begin{equation}\label{eq_freq_vector}
		\begin{aligned}
			\nu _{i}^\mathrm{freq}={p_{\max}}^{- \left(2\times i\right)/ c},
		\end{aligned}
	\end{equation}
	where $c$ is the channel dimension, and $i\in \left[ 0,1,\cdots ,c/2-1 \right] $. Meanwhile, we normalize $\bm{\beta}$ and get
	\begin{equation}\label{norm_beta}
		\begin{aligned}
			\boldsymbol{\beta }_{\mathrm{norm}}=\boldsymbol{\beta }/\beta _{\max}\times p_{\max}.
		\end{aligned}
	\end{equation}
	After that, we calculate the sine and cosine components of sinusoidal embedding features, i.e.,
	\begin{equation}\label{eq_feature_cos_sin}
		\begin{aligned}
			&\boldsymbol{g}_{\sin}=\sin\left( \boldsymbol{\beta }_{\mathrm{norm}}\odot \boldsymbol{\nu }^{\mathrm{freq}} \right) , \\ 
			&\boldsymbol{g}_{\cos}=\cos\left( \boldsymbol{\beta }_{\mathrm{norm}}\odot \boldsymbol{\nu }^{\mathrm{freq}} \right),
		\end{aligned}
	\end{equation}
	where $\odot$ denotes element-wise product which is operated on the broadcasted dimensions $(c/2)\times w \times h$. Finally, the global features $\bm{g} \in \mathbb{R}^{c \times w \times h}$ are expressed as 
	\begin{equation}\label{eq_feature_global}
		\begin{aligned}
			\boldsymbol{g}=\mathrm{MLP}\left( \mathrm{concatenate}\left[ \boldsymbol{g}_{\sin}, \boldsymbol{g}_{\cos} \right] \right),
		\end{aligned}
	\end{equation}
	where MLP consists of two linear layers with $c$ channels and ReLU activations, and $\mathrm{concatenate}$ denotes a concatenation operation between two features. To match multi-scale features of $g_s$, $\bm{g}$ are then up-sampled to $\bm{g}_1 \in \mathbb{R}^{c \times 2w \times 2h} $, $\bm{g}_2 \in \mathbb{R}^{c \times 4w \times 4h} $, and $\bm{g}_3 \in \mathbb{R}^{c \times 8w \times 8h}$.
	
	For the generator $g_s$, we replace its residual block with a conditioning residual bottleneck (Cond RB). In each Cond RB, we first align the corresponding global feature $\bm{g}_i$ to match the channel dimensions of the intermediate features output by each convolutional layer (each uses a linear layer), and then add them together. We note that we only add a lightweight control module at the receiver, which allows us to decode reconstructions for any realism map $\bm{\beta}$ from a single received signal in a one-shot transmission.
	
	\subsection{Spatial-Weighted Training Strategy}\label{Model Training Strategy}
	To train our DPCT model, we develop a spatial-weighted training strategy, where the optimization objective can adjust the perception penalty at different locations according to the realism map $\bm{\beta}$. Specifically, We first pre-train a model without SREM, optimized for the RD trade-off using \eqref{loss_RD_func}. Based on the pre-trained model, we optimize the RDP trade-off by \eqref{loss_F_DPCT} and \eqref{loss_D} for the full DPCT model. Unlike a single value in \eqref{loss_F_DPCT}, the weight of the perception term in \eqref{loss_F_DPCT} is a vector map. Therefore, the GAN loss should be weighted spatially. Our discriminator $D$ is well-suited to meet this requirement: it assigns penalties on image patches, essentially assessing whether each patch in the reconstructed image is fake or real. This means that the output of the discriminator, i.e., $D\left( \boldsymbol{x},\boldsymbol{y} \right) \in \mathbb{R}^{w \times h} $, is also spatially aligned with the image. Thus, \eqref{loss_F} is reformulated as
	\begin{equation}\label{loss_F_DPCT}
		\begin{aligned}
			&\mathcal{L} _{\mathcal{F}}^{\prime}=\mathbb{E} _{\boldsymbol{x}\sim p_{\boldsymbol{x}}} \mathbb{E} _{\boldsymbol{\beta}\sim p_{\boldsymbol{\beta}}}\Big( \underbrace{ \mathrm{MSE}\left( \boldsymbol{x},\boldsymbol{\hat{x}} \right)}_{\text{distortion metric}} +\lambda \underbrace{ \left( -\eta \log p_{\boldsymbol{y}}\left( \boldsymbol{y} \right) \right)}_{\text{transmission rate}} \\ &\,\,\,\,\,\,\,\,\,\,\,\,\,\,\,\,  +\mathcal{A}  \big(\underbrace{ \bm{\beta} \odot \left( -\log D\left( \boldsymbol{\hat{x}},\boldsymbol{y} \right) + C_P\mathcal{L} _P \left( \boldsymbol{x},\boldsymbol{\hat{x}} \right)	\right)}_{\text{spatial-weighted perception metric}} \big) \Big),
		\end{aligned}
	\end{equation}
	where we use the MSE distortion metric and add the LPIPS loss $\mathcal{L} _P$ (a ``perceptual'' distortion metric based on networks, used for stabilizing training) into the perception metrics, and $C_P$ is a hyper-parameter to weight the LPIPS loss and the GAN loss. We emphasize that LPIPS is also configured to be spatial. The function $\mathcal{A} \left( \cdot \right) $ denotes average pooling calculation. During the training phase, we first set the realism map to a constant matrix for the first 80\% training steps, i.e., $\bm{\beta }=b\cdot \mathbf{J}$, where $b$ is randomly sampled from $\left[ 0, \beta _{\max} \right]$. For the remaining 20\% steps, every element $\beta_{i,j}$ of $\bm{\beta}$ is independently sampled from $\left[ 0, \beta _{\max} \right]$.
	
	\section{CCT: Content-Aware Control for DP Trade-off under Scarce Bandwidth}\label{CCT}
	In scenarios with scarce bandwidth, RD-optimized methods can not only lose all textures but also suffer from content loss. To address this, beyond optimizing global appearance realism with DPCT, we further explore the generative capabilities of our framework to enhance content consistency under bandwidth constraints.
	
	When user is interested in specific regions of an image, it becomes unnecessary to transmit other regions in low distortion, allowing us to generate them completely to reduce channel bandwidth costs. In this case, the GAN must generate \emph{realistic textures} for entire image and \emph{consistent content} for untransmitted regions to ensure a global satisfactory human perception.
	To achieve this, we leverage the biased property of human perception within our RDP-JSCC framework to develop a new model, called content-controllable transmission (CCT). CCT is designed to completely generate user-uninteresting regions in the image from an instance label map while other regions are preserved with lossless contents. This reduces bandwidth cost proportionally, as fully generated regions do not require any coding stream to transmit them. Besides, extensive textures are generated over the entire image like DPCT; even though textures of the fully generated regions deviate more from the original, they can still serve the intended purpose. For example, in video calls, the synthesized background does not affect user perception. Furthermore, through user prompt feedback from the receiver to the sender, our CCT can support an interactive and scalable JSCC system to progressively satisfy the user perception needs.
	
	\begin{figure*}[t]
		\setlength{\abovecaptionskip}{0cm}
		\setlength{\belowcaptionskip}{0cm}
		\centering{\includegraphics[scale=0.4]{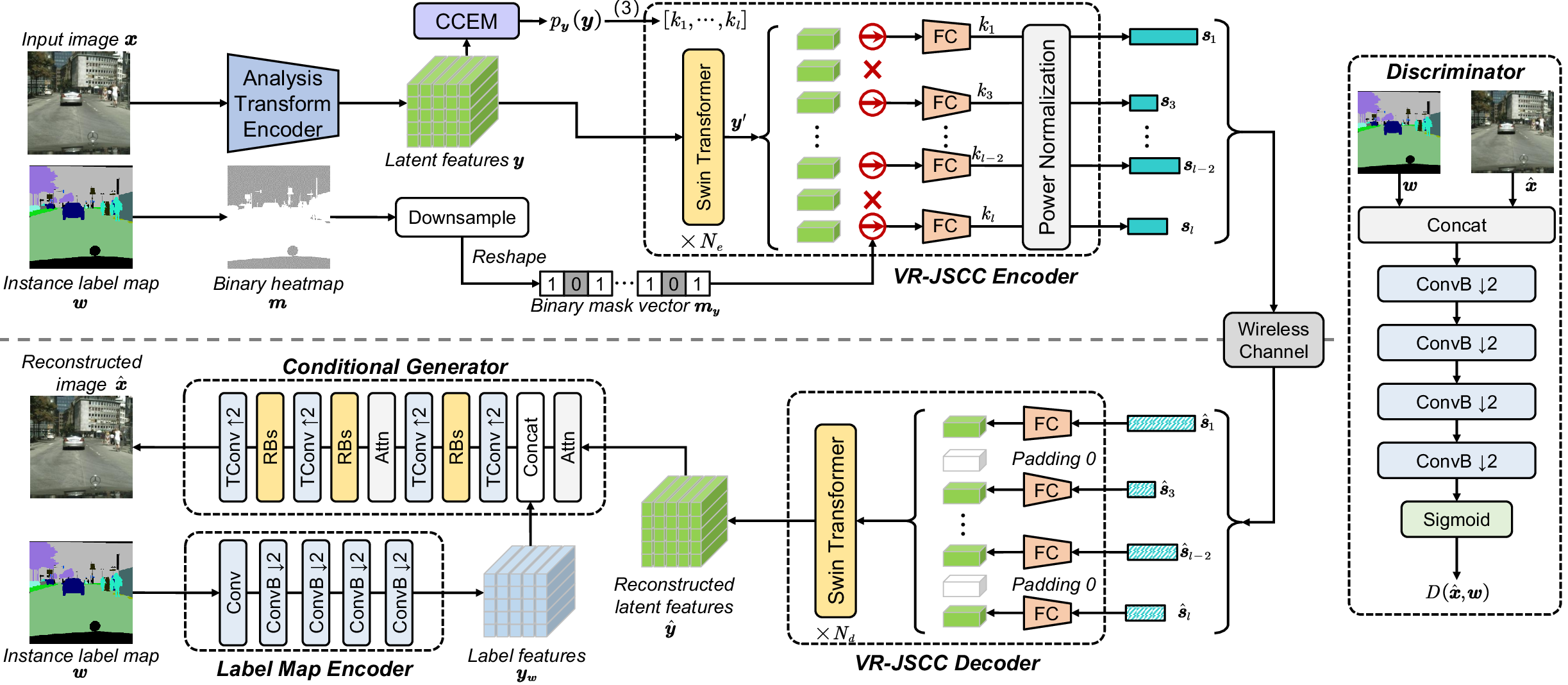}}
		\caption{The network architectures of the CCT model. We present the details of VR-JSCC to show how each embedding is guided by the instance label map $\bm{w}$, and function \eqref{eq_channel_bandwidth_cost_cal} calculates the bandwidth cost for each embedding. Both the generator and the discriminator $D$ condition on $\bm{w}$, A label map encoder is used for conditioning the generator on $\bm{w}$, and the discriminator $D$ is also conditioned on $\bm{w}$. \emph{ConvB} represents a convolutional block, each includes one convolutional layer, one normalization layer, and one activation function. The instance label map is transmitted as a side information through the reliable coded transmission link (e.g. 5G digital link) to the receiver, and it takes about 3.3\%$\sim$4.5\% of the total bandwidth cost for transmitting $1024\times2048$ resolution images in our CCT setting.}
		\label{fig4}
		\vspace{-1em}
	\end{figure*}
	
	\subsection{Content-Aware Selective Transmission Paradigm}\label{Label_map_driven_content_content}
	As shown in Fig. \ref{fig4}, the transmitter only transmits the coding stream of certain regions of the image $\bm{x} \in \mathbb{R}^{3\times W\times H}$. This is specified by a pixel-level binary heatmap $\bm{m} \in \mathbb{R}^{1\times W\times H}$, where ``1'' in $\bm{m}$ denotes that the corresponding image regions will be transmitted, while the pixel of ``0'' denotes untransmitted regions. Additionally, a pixel-level instance label map $\bm{w} \in \mathbb{R}^{3\times W\times H}$ of $\bm{x}$ is available at the transceiver, with pixels of the same RGB value in $\bm{w}$ belonging to the same instance, such as trees, sky, or people. We down-sample $\bm{m}$ to match the spatial dimensions of $\bm{y}$ and reshape it to a binary mask vector $\bm{m}_{\bm{y}} \in \mathbb{R}^{1\times l}$, where $l=w\times h$. Subsequently, we use $\bm{m}_{\bm{y}}$ to directly guide which embeddings in $\bm{y}{\prime}$ (the output of swin transformer) should be transmitted and which should be dropped.
	
	At the receiver, the noise channel vectors $\bm{\hat{s}}_i$ will be resized to the same dimension, then combined with 0-padding embeddings to recover the latent features $\hat{\bm{y}}$ using JSCC decoder $f_d$. Next, we take $\bm{w}$ as the conditional information to provide the location knowledge for the generator $g_s$. Instead of being fed directly into $g_s$, $\bm{w}$ is sent to a label map encoder $g_l$ to extract the label features $\bm{y}_{\bm{w}}$, which is then concatenated with intermediate features of $g_s$. During training, we also take $\bm{w}$ as conditional information for the discriminator $D$. Accordingly, the formulation of conditional GANs in the CCT model can be expressed as 
	\begin{equation}\label{eq_G_D_refine_CCT}
		\begin{aligned}
			G\leftarrow g_s\left( \boldsymbol{\hat{y}}, g_l\left( \boldsymbol{w} \right) \right) , D\leftarrow D\left( \boldsymbol{\hat{x}},\boldsymbol{w} \right).
		\end{aligned}
	\end{equation}
	Notably, $\bm{w}$ can be obtained through readily available segmentation models (such as SAM \cite{kirillov2023segment}) at the transmitter, stored as a simple vector graphic, and transmitted via a reliable coded transmission link (e.g., 5G coded transmission \cite{richardson2018design}). This process consumes minimal additional bandwidth, independent of the image dimension. Thus, the bandwidth cost of $\bm{w}$ as a percentage of total bandwidth cost decreases as image dimensions increase. For example, it constitutes only 3.3\%$\sim$4.5\% of the total bandwidth cost for transmitting $1024\times2048$ resolution images in our CCT setting. When the image resolution increases to $2048\times4096$ pixels, it is only 0.8\%$\sim$1.2\%.
	
	\subsection{Content-Aware VR-JSCC Codec Implementation}\label{masked_fe_fd_CCT}
	In this subsection, we mainly present the details of our content-aware VR-JSCC codec. Unlike the VR-JSCC in \cite{dai2022nonlinear}, the bandwidth cost for each embedding is determined not only by the entropy $\boldsymbol{p}_{\boldsymbol{y}}\left( \boldsymbol{y} \right)$ but also controlled by binary mask vector $\bm{m}_{\bm{y}}$ to further save bandwidth. In other words, our content-aware VR-JSCC can transmit the contents specified by the user.
	
	Specifically, given $\boldsymbol{p}_{\boldsymbol{y}}\left( \boldsymbol{y} \right)$, the function \eqref{eq_channel_bandwidth_cost_cal} calculates the bandwidth cost for each embedding, yielding the rate allocation vector $\bm{k}=\left[ k_1,\cdots ,k_l \right] $. Besides, we impose an additional control: using $\bm{m}_{\bm{y}}$ to control each embedding whether to be transmitted. Guided by $\bm{k}$ and $\bm{m}_{\bm{y}}$, $f_e$ realizes a variable-rate transmission for transmitting the latent representation $\bm{y}$, and $f_e$ incorporates a robust shared backbone based on $N_e$ swin transformer blocks to learn the contextual dependencies among $\bm{y}_i$. Meanwhile, we institute a collection of learnable rate tokens, $\mathcal{R} =\left \{r_0, r_{v_1},r_{v_2},\cdots ,r_{v_{2^q}} \right \}$, to convey the CBR information, where $r_0$ is for untransmitted embeddings. After added with $\mathcal{R}$, $\bm{y}$ is fed into swin transformer blocks to obtain $\bm{y}\prime$. Additionally, light FC layers are utilized to compress each $\bm{y}_i^{\prime}$ into a $k_i$-dimensional channel symbol vector $s_i$. We utilize a set of $2^q$ FC layers, each having unique output dimensions $\{ v_1, v_2, \dots, v_{2^{q}} \}$, activated as needed.
	
	At the receiver, FC blocks reshape noisy channel vectors $\hat{s}_i$ with various lengths to the same dimension, and missing positions are replaced by 0-padding embedding. Then they are added to the rate tokens and fed in parallel into $N_d$ shared Transformer blocks to recover $\bm{\hat{y}}$. For other modules in Fig. \ref{fig4}, We show the details of the generator $g_s$ and the discriminator $D$, both of which are conditioned on $\bm{w}$. 
	
	\subsection{Masked Training Strategy}\label{Model_Training_Strategy_CCT}
	To optimize the CCT model, we adopt a masked training strategy, which aims to train the model to adapt well to any combination of transmitted instances. Particularly, we first pre-train an RD-optimized CCT model without any masking. In the next phase, we use a fixed weight factor $\beta$ to bias the model towards perception. Meanwhile, we randomly select only $25\%$ of all instances to transmit but masking others at every training step. After that, we can obtain a masked binary heatmap $\bm{m}$ to mask the MSE distortion (pixel-wise), which makes our model focus on improving the distortion of preserved instances while generating untransmitted instances. Therefore \eqref{loss_F} is reformulated as
	\begin{equation}\label{loss_F_CCT}
		\begin{aligned}
			&\mathcal{L} _{\mathcal{F}}^{\prime \prime}=\mathbb{E} _{\boldsymbol{x}\sim p_{\boldsymbol{x}}}\mathbb{E} _{\boldsymbol{m}\sim p_{\boldsymbol{m}}}\Big( \bm{m} \cdot \mathrm{MSE}\left( \boldsymbol{x},\boldsymbol{\hat{x}} \right) +\lambda R\\ 
			&\,\,\,\,\,\,\,\,\,\,\,\,\,\,\,\, +\beta \left( -\log D\left( \boldsymbol{\hat{x}},\boldsymbol{w} \right) + C_P\mathcal{L} _P \left( \boldsymbol{x},\boldsymbol{\hat{x}} \right)	\right) \Big),
		\end{aligned}
	\end{equation}
	where both $-\log D\left( \boldsymbol{\hat{x}},\boldsymbol{w} \right)$ and $\mathcal{L} _P \left( \boldsymbol{x},\boldsymbol{\hat{x}} \right)$ are scalars after average pooling calculation, and $R=-\eta \log p_{\boldsymbol{y}}\left( \boldsymbol{y} \right)$. Overall, our masked training strategy draws on the advantages of mask learning \cite{he2022masked}, significantly enhancing the representational capacity and robustness of the entire model. Additionally, this strategy helps the transmission system proportionally save bandwidth cost as masked instances increase.
	
	\subsection{CCT Model Enabled Interactive and Scalable JSCC }\label{Interactive_and_Scalable_JSCC}
\begin{figure}[t]
		\setlength{\abovecaptionskip}{0cm}
		\setlength{\belowcaptionskip}{0cm}
		\centering{\includegraphics[scale=0.39]{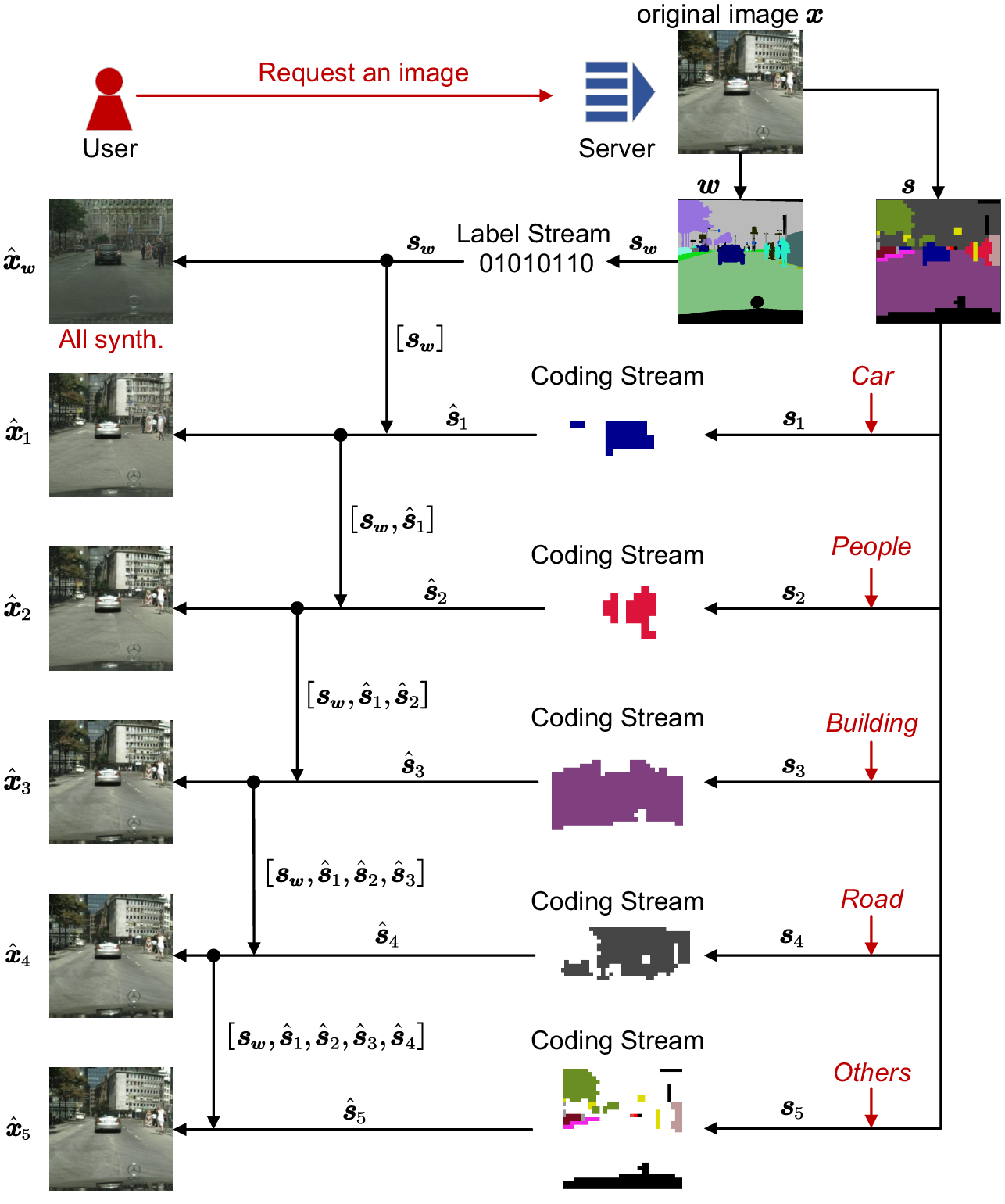}}
		\caption{Illustration about the process of an interactive and scalable JSCC system based on the CCT model.}
		\label{fig5}
		\vspace{-1em}
	\end{figure}

	Given a well-trained CCT model as above, we can further realize an interactive and scalable JSCC transmission system by introducing the user prompt feedback mechanism, as shown in Fig. \ref{fig5}.
	
	\subsubsection{User Prompt Feedback Mechanism}\label{User_Prompt_feedback}
	Here, we consider a more specialized scenario where the user (the receiver) can specify any contents to preserve or generate through a simple prompt feedback to the server (the transmitter). For simplicity, instance labels such as \emph{car}, \emph{road}, or \emph{trees} are used as prompts, and each image is segmented into different instances. In this setting, the binary heatmap $\bm{m}$ can be easily constructed according to the instance label map $\bm{w}$ and the label prompt, and the bandwidth cost used for sending prompts can be negligible, as the prompts are texts and easy to be compressed to an extremely tiny bandwidth cost. 
	
	\subsubsection{Interactive and Scalable Properties}\label{Interactive_and_Scalable}
	``Interactive'' means that multiple communications can be operated between the user and the server to progressively meet the user's needs. As shown in Fig. \ref{fig5}, the user first requests for an image, and the server then transmits the instance label map $\bm{w}$. The user can then generate a preliminary image from $\bm{w}$, which provides an approximation of the content but lacks the most realistic textures, i.e., it allows the user to know the approximate contents but is heavily damaged. Subsequently, the user sends prompts $P_i$ to make the specific instances clearer and sharper (realistic textures), while other areas are also generated to enhance the overall perceptual quality. If the desired subjective perception is not achieved, the user may continue to communicate until satisfied.
	
	\begin{figure*}[t]
		\setlength{\abovecaptionskip}{0cm}
		\setlength{\belowcaptionskip}{0cm}
		\centering{\includegraphics[scale=0.5]{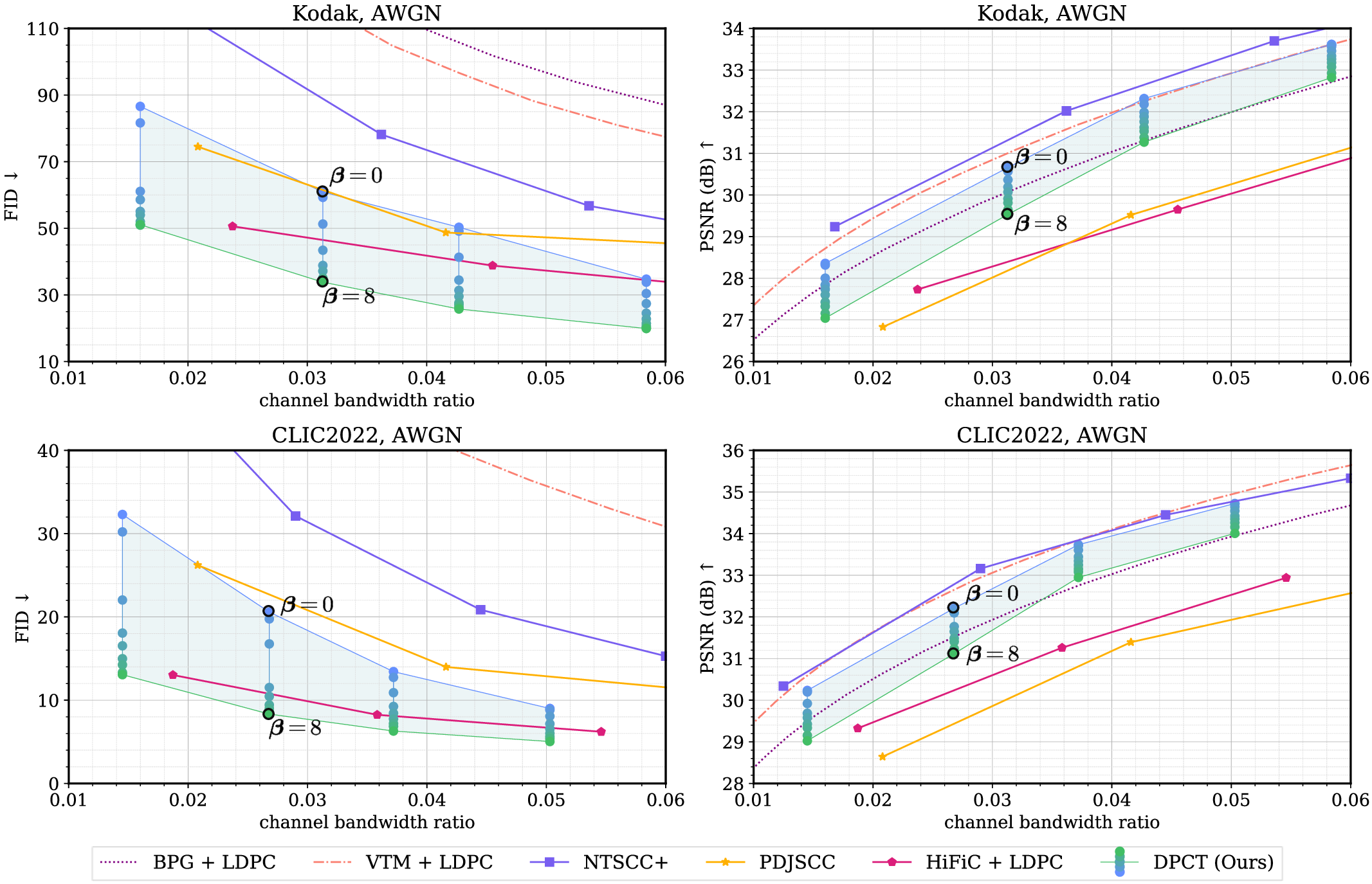}}
		\caption{CBR-FID curves and CBR-PSNR curves on Kodak (the first row) and CLIC2022 (the second row). FID is a measure of \emph{perception} ($\downarrow$ indicates lower is better) and PSNR is a measure of \emph{distortion} ($\uparrow$ indicates higher is better). All experiments are taken over the AWGN channel at SNR = $10$dB. For our DPCT model, we decode multiple distortion-perception points by varying $\bm{\beta}$ from $0$ to $8$ on the receiver side, using a single model per CBR.}
		\label{fig6}
		\vspace{0em}
	\end{figure*}
	
	Before responding to the first prompts, the server outputs the whole sequence of channel symbols $\bm{s}$ without masking, then divides them into multiple transmission streams $\left[ \boldsymbol{s}_1,\boldsymbol{s}_2,\cdots ,\boldsymbol{s}_n \right]$ corresponding to different instances. All streams are stored in the cache. In response to $P_i\left(  i= 1,2,\dots \right) $, the server retrieves $\boldsymbol{s}_i$ from the cache and transmits it, and the user end reconstructs the image $\hat{\bm{x}}_i$ from $\left[ \boldsymbol{w},\hat{\bm{s}}_1,\cdots ,\hat{\bm{s}}_i \right]$. This agrees with the ``scalable'' concept, where we first transmit only the label stream $\bm{s}_{\bm{w}}$, and whether to pass coding streams $\bm{s}_i$ depends on the request for ever-increasing instances of the user. The scalability of our CCT is facilitated by the two facts that: (1) $\bm{s}$ is divisible in spatial dimensions; (2) the user end can reconstruct an image from the combination of the label stream $\bm{s}_{\bm{w}}$ and any noisy coding streams $\hat{\bm{s}}_i$.
	
	\section{Experimental Results}\label{Experimental_Results}
	\subsection{Experimental Setup}\label{Experimental_Setup}
	\subsubsection{Datasets}\label{Datasets}
	We train DPCT models for transmission using 500,000 diverse natural images sampled from the Open Images dataset \cite{OpenImage} and we use the medium-size Kodak dataset \cite{Kodak} (24 images, $768 \times 512$ pixels) as well as the large-size CLIC2022 dataset \cite{CLIC2022} (including validation set and test set, 60 images, up to $2048 \times 2048$ pixels) for testing. During DPCT model training, images are randomly cropped into $256\times256$ patches. CCT models are trained on the Cityscapes dataset \cite{cordts2016cityscapes} (2975 training images, validation 500 images, $2048 \times 1024$ pixels). We randomly select 20 validation images to evaluate them and resize both training and validation images to $1024 \times 512$ pixels.
	
	\subsubsection{Comparison Schemes}\label{Comparison_Schemes}
	To evaluate our proposed methods, we conducted comparative analyses with established transmission techniques across several categories. For separation-based
	image transmission schemes, our benchmarks include an array of mainstream image codecs, both handcrafted and learned, paired with 5G LDPC channel coding \cite{richardson2018design} (code length 4096) and digital modulation in line with the 3GPP TS 38.214 standard. Specifically, the handcrafted image codecs in consideration encompass BPG \cite{BPG} (compliant with the intra-frame coding scheme of the HEVC standard), and VTM \cite{bross2021developments} (intra-frame coding scheme of the VVC standard, SOTA engineered image codec). The learned image codec assessed is HiFiC \cite{mentzer2020high} (a former long-term SOTA in terms of RDP-optimized NICs). These three separation-based schemes are marked as ``BPG + LDPC'', ``VTM + LDPC'', and ``HiFiC + LDPC''. In the category of end-to-end JSCC methods, we consider \emph{NTSCC+}, the latest version of NTSCC, proposed in \cite{dai2022nonlinear}, and \emph{PDJSCC}, the first JSCC work for human perception under high-resolution images. 
	
	\subsubsection{Evaluation Metircs and User Study}\label{Evaluation_Metircs}
	To evaluate the distortion performance, we calculate the peak signal-to-noise ratio (PSNR) of each pair of the original and reconstructed images. To evaluate the perception performance, we use FID \cite{heusel2017gans}, which has been widely used in various tasks to assess human perceptual quality. Due to the large variation of resolutions in our validation sets, we calculate FID on image patches rather than on full images. We use a patch size of $128$ pixels for Kodak resulting in $936$ patches, and $256$ pixels for CLIC2022 resulting in $4290$ patches. In addition, for the CCT model, we also adopt the mean IoU \cite{csurka2013good} to evaluate the semantic consistency. We use ``$\downarrow$'' and ``$\uparrow$'' to mark that lower or higher metrics represent better quality.
	
	For a more subjective assessment on CCT, we use the ``two alternatives, forced choice'' test for quantitative assessment in our user study. The user study system we design shows human raters an interactive window, where the raters can drag the slider to compare the original image with one produced from our methods or ``VTM + LDPC'' method. We require raters to vote for the image which looks visually more pleasing. We randomly pick 5 images from Cityscapes validation images. 
	
	\begin{figure*}[t]
		\setlength{\abovecaptionskip}{0cm}
		\setlength{\belowcaptionskip}{0cm}
		\centering{\includegraphics[scale=0.44]{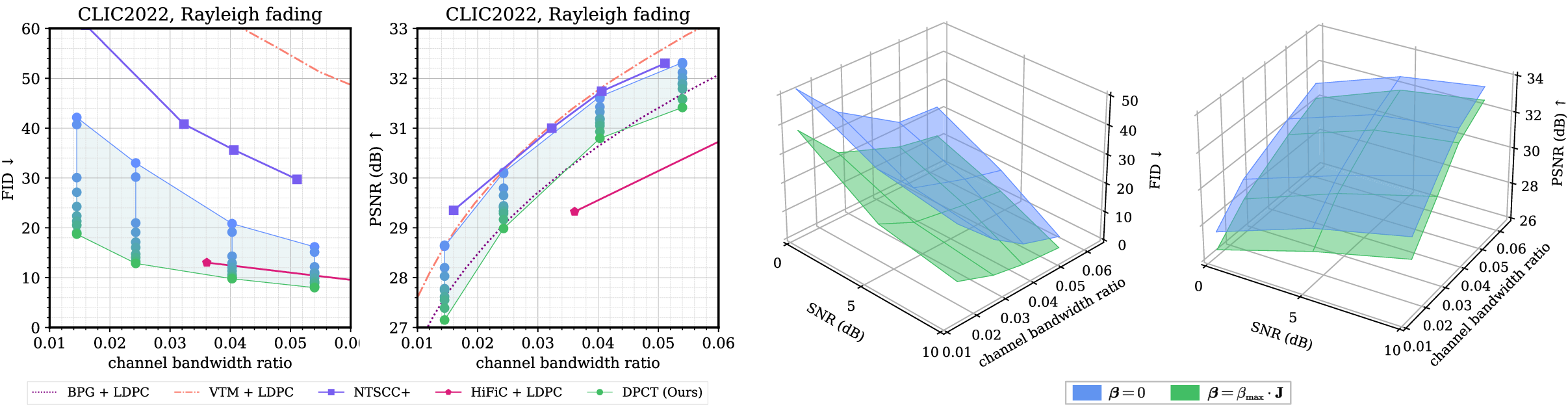}}
		\caption{Results under variable channel states. Left two: CBR-FID performance and CBR-PSNR performance on CLIC2022 over the Rayleigh fading channel at average SNR $=10$dB. Right two: SNR-CBR-PSNR performance and SNR-CBR-FID performance on CLIC2022 over the AWGN channel. For the digital systems, we adjust the channel coding rate and modulation order to ensure reliable transmission and the highest efficiency under different channel states.}
		\label{fig7}
		\vspace{-1em}
	\end{figure*}
	
	\begin{figure}[t]
		\setlength{\abovecaptionskip}{0cm}
		\setlength{\belowcaptionskip}{0cm}
		\centering{\includegraphics[scale=0.5]{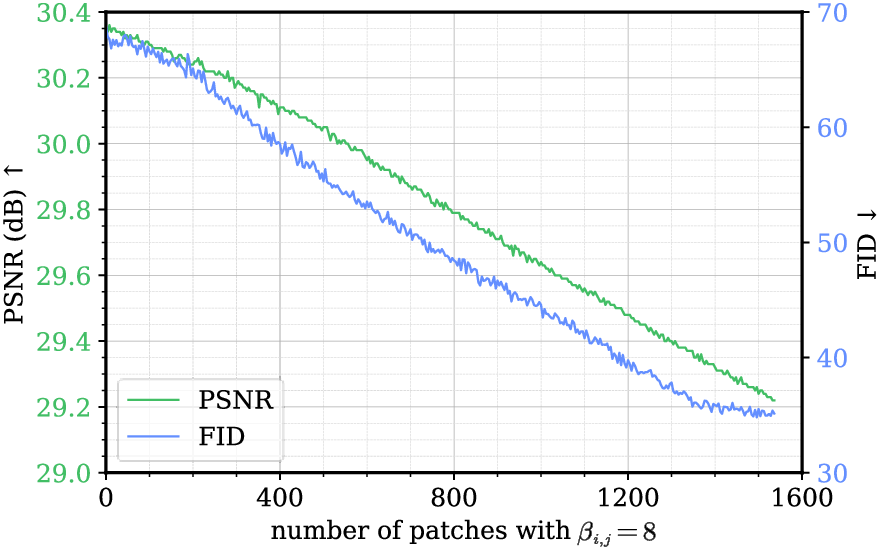}}
		\caption{The fine-grained spatial control of our DPCT over the DP trade-off. The FID (right Y-axis) and PSNR (left Y-axis) results vary with the number of patches with $\beta_{i,j}=8$ on Kodak. We initialize the realism map to $\bm{\beta}=0$ and then incrementally set $\beta_{i,j}=8$ for each $16\times16$ image patch in sequential rows until $\bm{\beta}=8$. For each Kodak image, the size of $\bm{\beta}$ is $32\times48$ or $48\times32$, and there are $1536$ patches in total.}
		\label{fig8}
		\vspace{-1em}
	\end{figure}
	
	\begin{figure}[t]
		\setlength{\abovecaptionskip}{0cm}
		\setlength{\belowcaptionskip}{0cm}
		\centering{\includegraphics[scale=0.65]{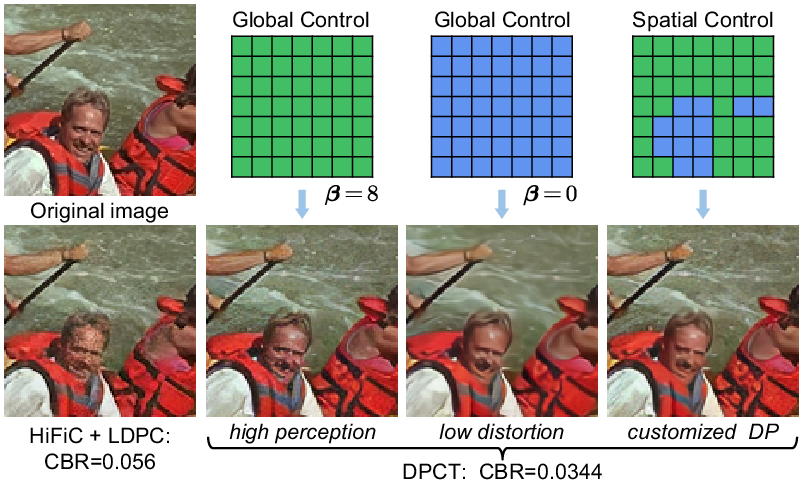}}
		\caption{A visual demo of using the spatial control mode to improve the visual quality of small faces. Both ``HiFiC + LPDC'' and DPCT ($\bm{\beta}=8$) generate rich details of the waves, but both over-distort the face, making it not so realistic. However, we can find that DPCT ($\bm{\beta}=0$) recovers the face faithfully, but it can not generate details for waves. Therefore, we set $\beta_{i,j}=0$ at the region of face and $\beta_{i,j}=8$ at the others to make the face more faithful, while generating realistic details for other regions. We note that the three images of DPCT are all decoded from the same $\bm{\hat{y}}$.}
		\label{fig9}
		\vspace{-1em}
	\end{figure}
	
	\begin{figure*}[t]
		\setlength{\abovecaptionskip}{0cm}
		\setlength{\belowcaptionskip}{0cm}
		\centering{\includegraphics[scale=0.455]{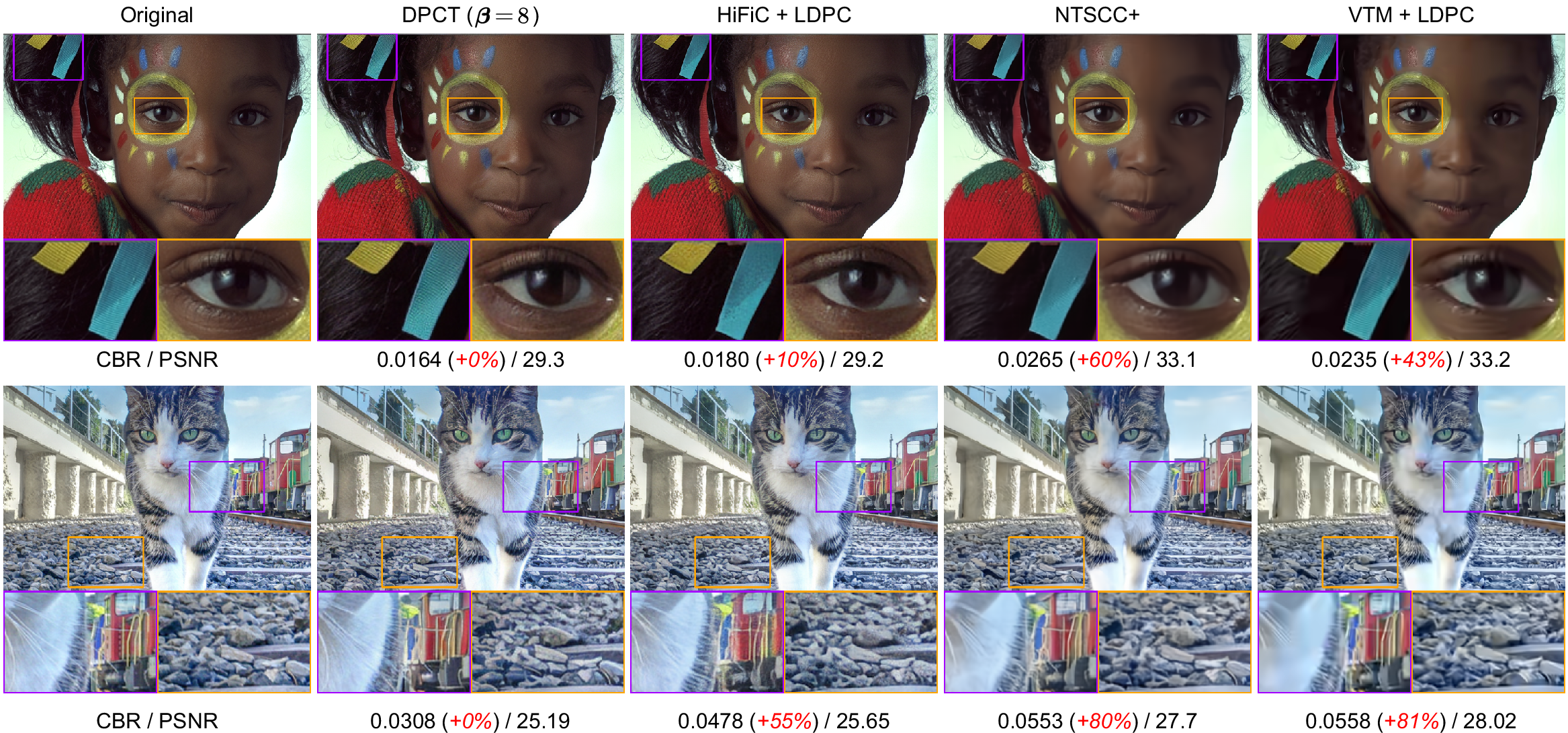}}
		\caption{Visual comparison between the input images between the reconstructions from our DPCT ($\bm{\beta}=8$), ``HiFiC + LDPC'', and NTSCC+. Images are transmitted over the AWGN channel (first row) and the Rayleigh channel (second row) at SNR = $10$dB. For each image, we present two crops of it for detail comparison. Red numbers indicate the percentage of bandwidth cost increase compared to our DPCT ($\bm{\beta}=8$). Note that our models always have the smallest CBR. Overall, we show how our high-realism reconstructions closely match the input, more than the other two. In the first row (AWGN), we can see the stripes on the headband, finer details in the hair, and more texture around the eyes, which ``HiFiC + LDPC'' does not render well. In the first row (Rayleigh fading), even though ``HiFiC + LDPC'' uses 55\% more the CBR, the details such as cat fur, stones, and the train are still not as realistic as ours, and are less close to the input. For NTSCC+, all the reconstructions are blurrier and lack details, even using 60\% or 80\% more the CBR.}
		\label{fig10}
		\vspace{0em}
	\end{figure*}
	
	\subsubsection{Models Training Details}\label{Models_Training_Details}
	For both DPCT and CCT models, we draw on the ``ELIC'' architectures \cite{he2022elic} for $g_a$ and $g_s$, and each layer has $3$ (Cond) RBs with the bottleneck dimension $N=256$. Besides, the VR-JSCC encoder and decoder $f_e$ and $f_d$ are built on $4$ Transformer blocks, i.e. $N_e=N_d=4$. The channel dimension $c$ of the latent space is set to $320$.
	The quantization value set $\mathcal{V}$ consists of $26$ values, which are evenly distributed from $1$ to $320$. The resulting rate allocation vector $\bm{k}$ is reliably transmitted as additional side information via a digital link (details can be found in \cite{wang2023improved}), and the cost of this side information ranges from 1\%$\sim$7\% of the total bandwidth. 
	
	For DPCT, we train models with different CBRs by setting $\lambda=0.1, 0.025, 0.005, 0.0015$, and fixing $\mathcal{L} _P=1$ and $\mu=0.2$. We consider the AWGN channel at SNR=$0, 5, 10$dB and the Rayleigh fading channel at SNR=$10$dB. During the RDP training phase, we first set the realism map to a constant matrix for the first 80\% training steps, i.e., $\bm{\beta }=b\cdot \mathbf{J}$, where $b$ is randomly sampled from $\left[ 0, \beta _{\max} \right]$. For the remaining 20\% steps, every element $\beta_{i,j}$ is independently sampled from $\left[ 0, \beta _{\max} \right]$ ($\beta _{\max}=8$ for SNR=$10$dB). Even though different $\lambda$s make different dynamics within equation \eqref{loss_F_DPCT}, the DP trade-off depends only on $\bm{\beta}$. 
	
	For CCT, we train two models over the AWGN channel at SNR $=10$dB by setting $\lambda=0.1, 0.025$, $\mathcal{L} _P=1$, and $\mu=0.2$. However, we use a fixed weight factor $\beta=8$ in \eqref{loss_F_CCT}. During the training phase of DPCT and CCT, we adopt the Adam optimizer \cite{ADAM} with a learning rate of $1\times10^{-4}$ and decay it by a factor $10\times$ in the last 50\% steps. The batch size is set to $8$ for DPCT and $1$ for CCT. All the implementations were executed utilizing Pytorch.
	
	\subsection{Results of DPCT}\label{Results_Analysis_DPCT}
	In Fig. \ref{fig6}, we show the CBR-FID and CBR-PSNR curves over the AWGN channel at SNR = $10$dB on Kodak (the first row) and CLIC2022 (the second row), respectively. For the digital communication systems, after traversing given combinations of LDPC coded modulation schemes, we exploit a 2/3 rate (4096, 6144) LDPC code with 16-ary quadrature amplitude modulation (16QAM) to ensure reliable transmission and the highest efficiency at SNR = $10$dB. For our DPCT model, we decode multiple distortion-perception points by varying $\bm{\beta}$ from $0$ to $8$ on the receiver side. Our DPCT using the global control mode achieves a remarkable performance in terms of distortion-perception: on the perception side ($\bm{\beta}=8$), our models dominate compared to all other methods, while outperforming significantly the generative methods PDJSCC and ``HiFiC + LDPC'' in PSNR. On the distortion side ($\bm{\beta}=0$), we show strong PSNR, reaching towards the SOTA learned and handcrafted MSE end-to-end methods, i.e., NTSCC+ and ``VTM + LDPC'', while superior obviously to them in FID. We highlight that it means reconstructions by our DPCT models in the perception setting are much closer to the input than PDJSCC and ``HiFiC + LDPC'', and we also realize greater appearance realism than NTSCC+ and ``VTM + LDPC'' in the distortion setting.
	
	In Fig. \ref{fig7}, we provide additional results under various channel conditions. For the digital schemes, we adjust the channel coding rate and modulation order to ensure reliable transmission and the highest efficiency. In two leftmost subfigures, we evaluate all schemes over the Rayleigh fading channel at SNR=$10$dB, and we assume perfect channel estimation, thereby the receiver is able to obtain an accurate $\bm{h}$ and employs zero-forcing equalization. The results show that our DPCT model demonstrates competitiveness under the Rayleigh fading channel similar to that under the AWGN channel: its perception performance still surpasses all comparison schemes, and its distortion performance at $\bm{\beta}=0$ and $\bm{\beta}=\beta_{\max}$ approaches that of ``BPG + LDPC'' and ``VTM + LDPC'' respectively. In the two rightmost subfigures, we present the FID and PSNR results at a wide range of SNR and CBR over the AWGN channel. We can find that our DPCT models also work well at variable SNR: there is still a clear trade-off between the distortion setting ($\bm{\beta}=0$, blue face) and perception setting ($\bm{\beta}=\beta_{\max}\cdot \mathbf{J}$, green face), and any region between these two faces is also achievable by adjusting $\bm{\beta}$. 
	
	To demonstrate fine-grained spatial control over the DP trade-off, we present the FID and PSNR results as the number of patches with $\beta_{i,j}=8$ varies on Kodak in Fig. \ref{fig8}. Specifically, we initialize the realism map to $\bm{\beta}=0$ and then incrementally set $\beta_{i,j}=8$ for each $16\times16$ image patch in sequential rows until $\bm{\beta}=8$. The results show that as the number of patches with $\beta_{i,j}=8$ increases, PSNR gradually worsens while FID gradually improves. This indicates that our realism map $\bm{\beta}$ can precisely guide the generator to produce realistic textures in the corresponding regions, achieving patch-level control.
	
	Based on the patch-level control, we can use the realism map $\bm{\beta}$ to control the DP trade-off for different contents. This addresses a drawback of traditional perception-optimized models: generating excessive details for content regions to which the human eyes are more sensitive, causing visually noticeable discomfort, such as small faces (``HiFiC + LDPC'' and DPCT ($\bm{\beta}=8$) in Fig. \ref{fig9} overly distort faces). This issue arises from applying the same optimization strategy to all contents. To mitigate this, we set $\beta_{i,j}=0$ in the facial region and $\beta_{i,j}=8$ in other regions to make the face more faithful while generating reasonable textures for other contents. For small faces, low pixel consistency is more important \cite{li2022content}.
	
	To intuitively demonstrate the remarkable perception performance ($\bm{\beta}=8$) of our DPCT model, we give some examples of side-by-side visual comparison under the AWGN (first row) and Rayleigh fading (second row) channel in Fig. \ref{fig10}. Compared to ``HiFiC + LDPC'', the textures generated by our DPCT are not only more realistic but also closer to the input. For example, we reconstruct the parallel stripes on the headband instead of the meaningless spots produced by ``HiFiC + LDPC''. This advantage is even more pronounced on the Rayleigh fading channel: ``HiFiC + LDPC'' uses 50\% more than the CBR, yet its generated textures are still not as realistic as ours. As for NTSCC+, due to the lack of distribution optimization, it tends to yield pixel-wise averages of plausible solutions that are overly smooth and lack details, even when using 80\% more the CBR than ours.
	
	\begin{figure*}[t]
		\begin{center}
			\subfigure{\includegraphics[width=1\textwidth]{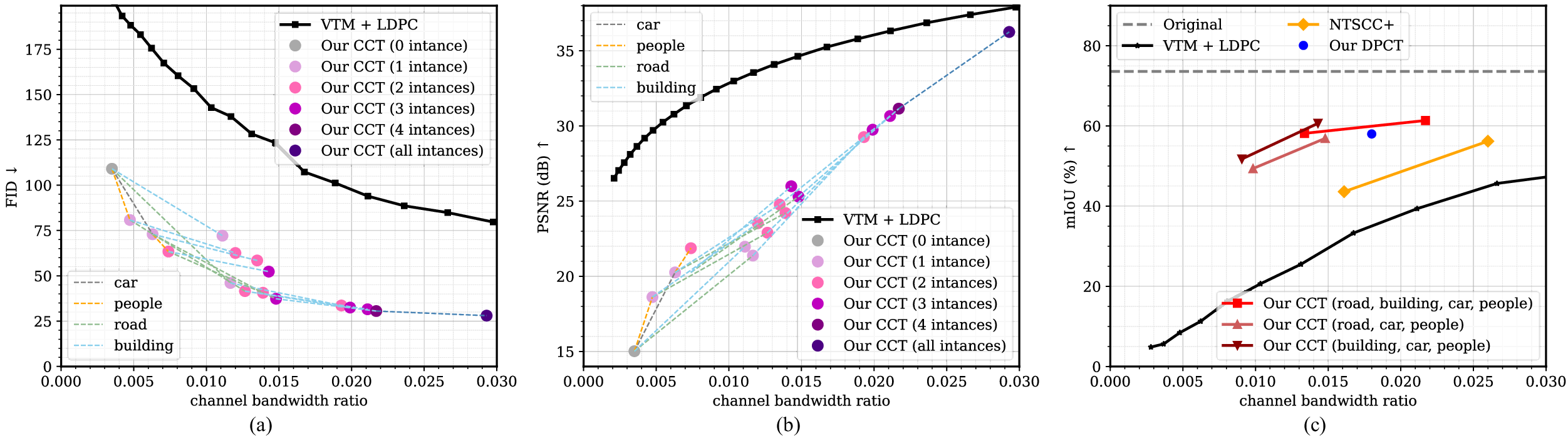}}
			\caption{(a) and (b): Evaluation for the interactive user-server process based on CCT, including the PSNR and FID performance on Cityscapes validation set at SNR = $10$dB. Each colorful data point denotes one transmission for one combination from \{\emph{car}, \emph{road}, \emph{people}, \emph{building}\}, and the dashed line connecting two data points denote the added instance between two combinations. The left-most data point corresponds to the transmission for instance label maps (CBR = $0.0035$ for $512\times1024$ resolution images). Inside the parentheses is the number of preserved instances. (c): Mean IoU in relation to CBR on Cityscapes dataset of CCT, DPCT ($\bm{\beta}$), NTSCC+, and ``VTM + LDPC'', and the gray dash line denotes the mIoU of the original images in Cityscapes validation set. We consider three situations for our CCT: (1) transmit \emph{road}, \emph{building}, \emph{car}, and \emph{people}; (2) transmit \emph{road}, \emph{car}, and \emph{people}; (3) transmit \emph{building}, \emph{car}, and \emph{people}. Others are generated.}
			\label{fig11}
			\vspace{-1em}
		\end{center}
	\end{figure*}
	
	\begin{figure}[t]
		\begin{center}
			\hspace{-.20in}
			\subfigure{\includegraphics[width=0.24\textwidth]{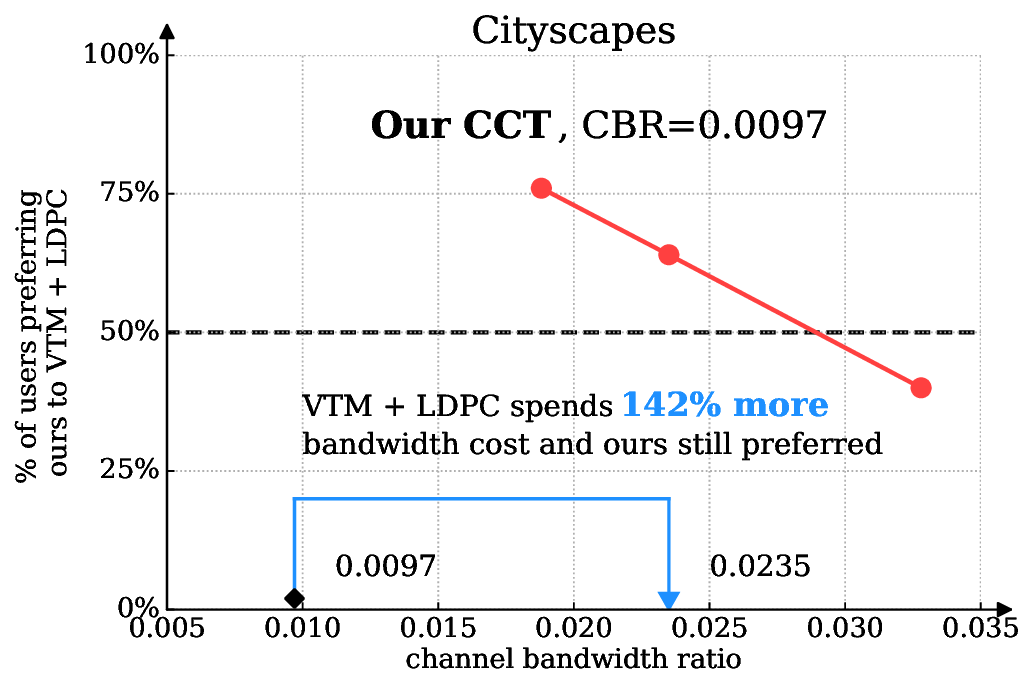}}
			\hspace{-.10in}
			\subfigure {\includegraphics[width=0.24\textwidth]{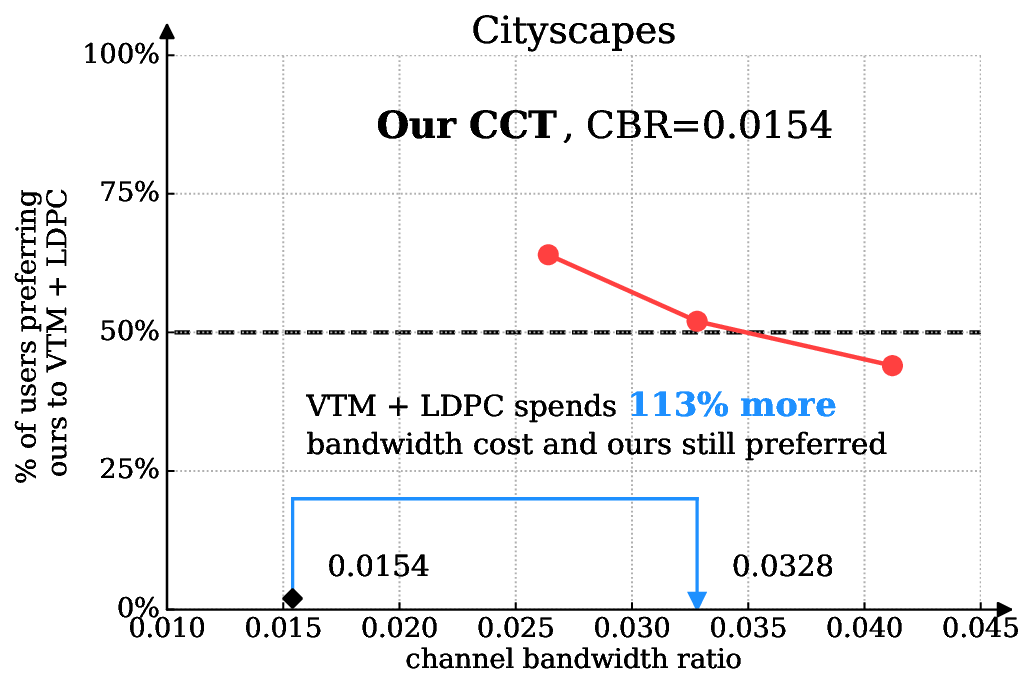}}
			\hspace{-.20in}
			\caption{The results of the user study for CCT, and the transmitted instances are fixed, including \emph{car}, \emph{people}, and \emph{building}.}
			\label{fig12}
		\end{center}
		\vspace{-2em}
	\end{figure}
	
	\begin{figure*}[t]
		\begin{center}
			\hspace{-.05in}
			\subfigure[CCT, all generated, 0.0034 (\textcolor{blue}{\textit{--90\%}})] {\includegraphics[width=0.244\textwidth]{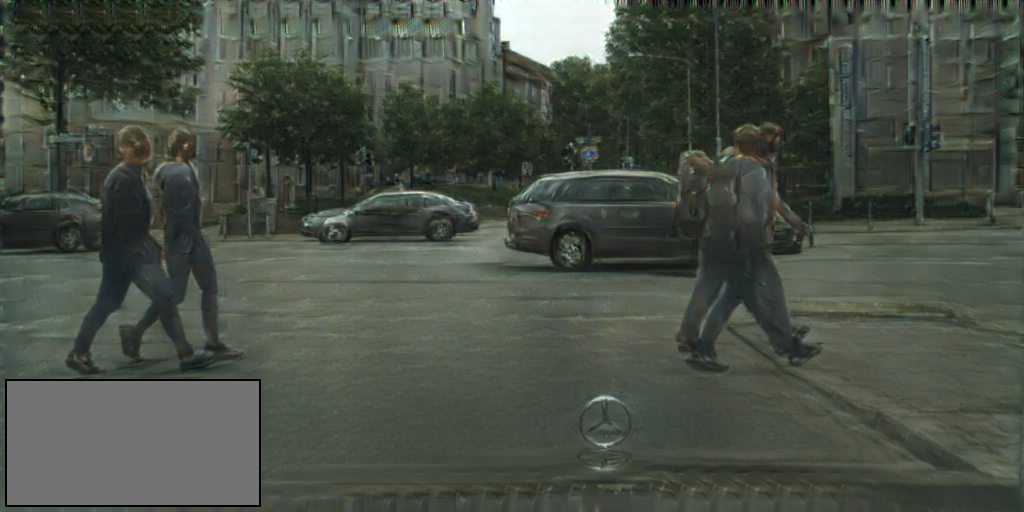}}
			\hspace{-.20in}
			\quad
			\subfigure[CCT, \emph{--car}, 0.0258 (\textcolor{blue}{\textit{--26\%}})] {\includegraphics[width=0.244\textwidth]{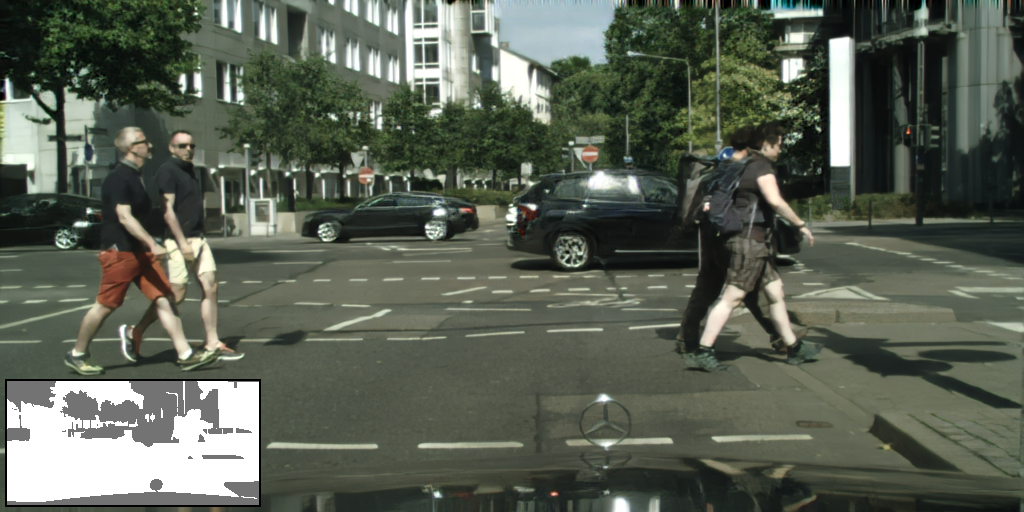}}
			\hspace{-.20in}
			\quad
			\subfigure[CCT, \emph{--road}, 0.0188 (\textcolor{blue}{\textit{--45\%}})] {\includegraphics[width=0.244\textwidth]{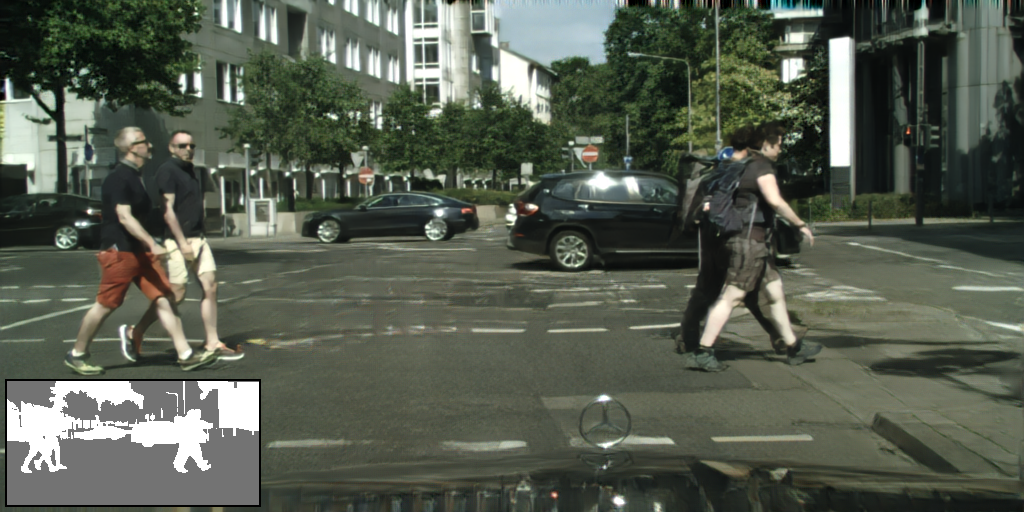}}
			\hspace{-.20in}
			\quad
			\subfigure[CCT, \emph{--building}, 0.0185 (\textcolor{blue}{\textit{--46\%}})] {\includegraphics[width=0.244\textwidth]{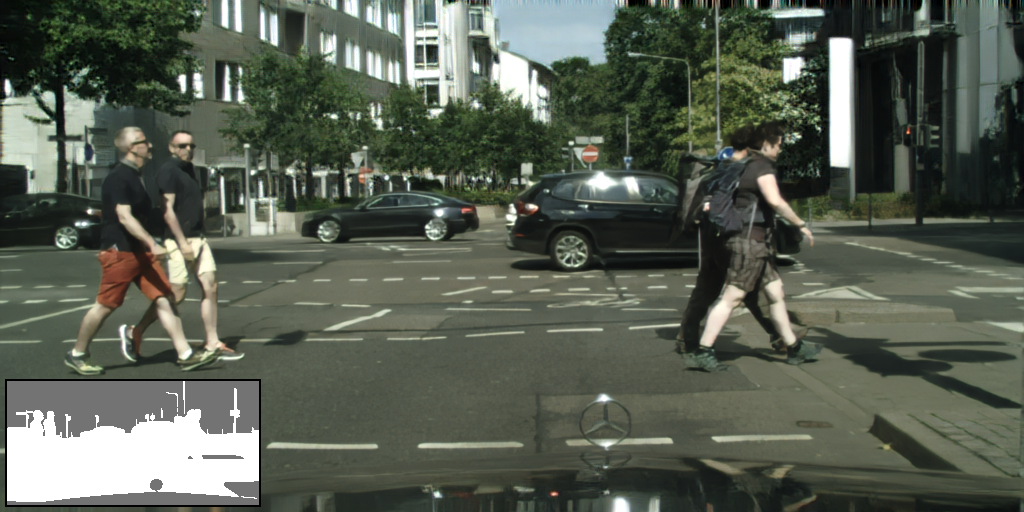}}
			
			\hspace{-.20in}
			\subfigure[CCT, \emph{--people}, 0.0256 (\textcolor{blue}{\textit{--25\%}})] {\includegraphics[width=0.244\textwidth]{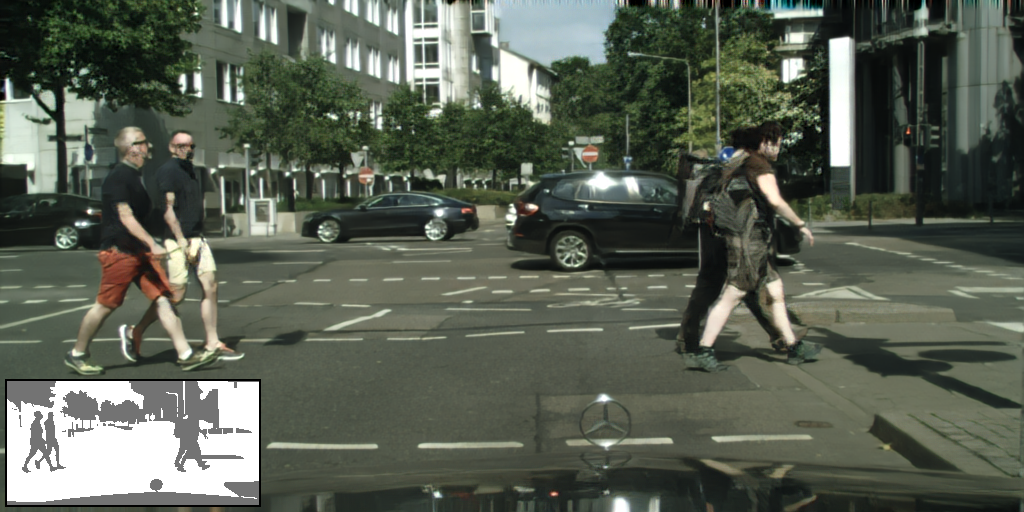}}
			\quad
			\hspace{-.20in}
			\subfigure[CCT, all trans., 0.0341 (\textcolor{black}{\textit{0\%}})] {\includegraphics[width=0.244\textwidth]{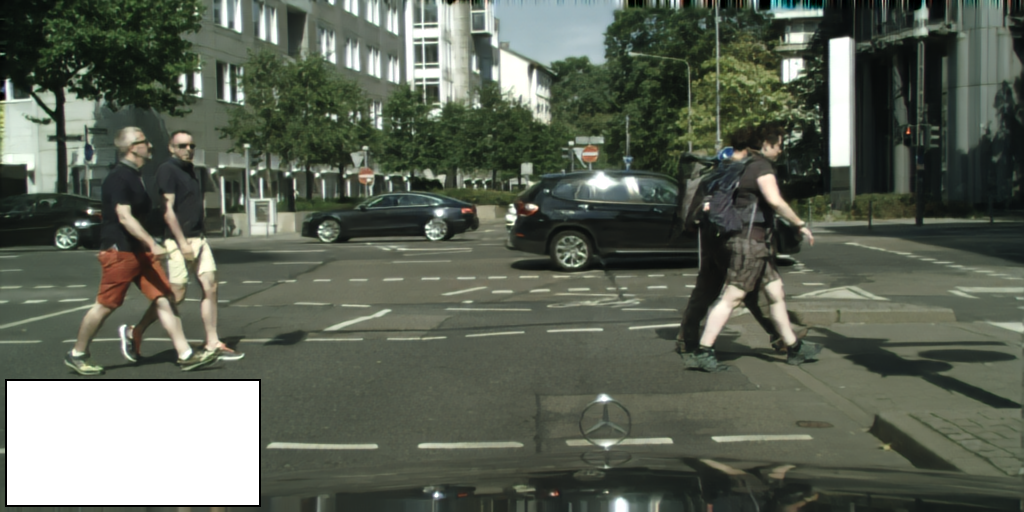}}
			\hspace{-.20in}
			\quad
			\subfigure[NTSCC+, 0.039 (\textcolor{red}{\textit{+14\%}})] {\includegraphics[width=0.244\textwidth]{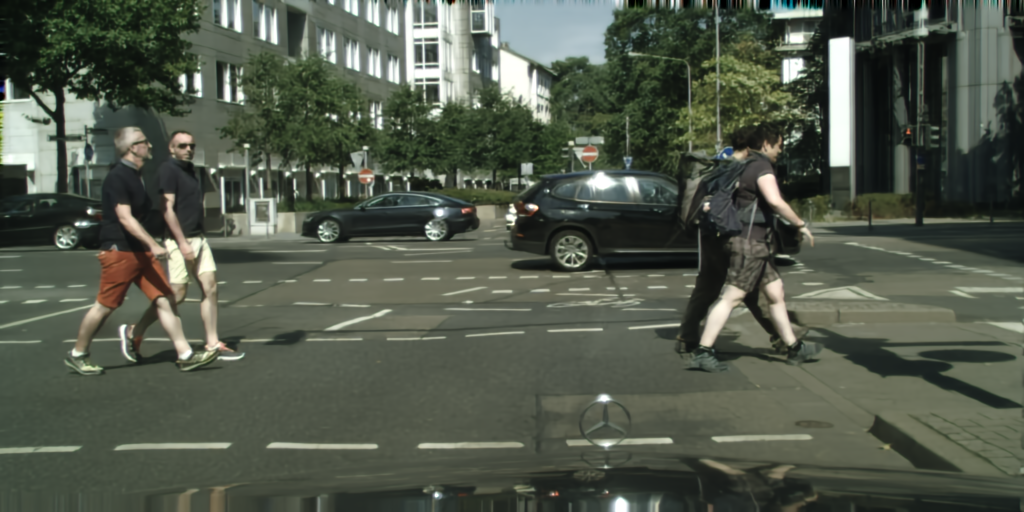}}
			\hspace{-.20in}
			\quad
			\subfigure[VTM + LDPC, 0.0404 (\textcolor{red}{\textit{+19\%}})] {\includegraphics[width=0.244\textwidth]{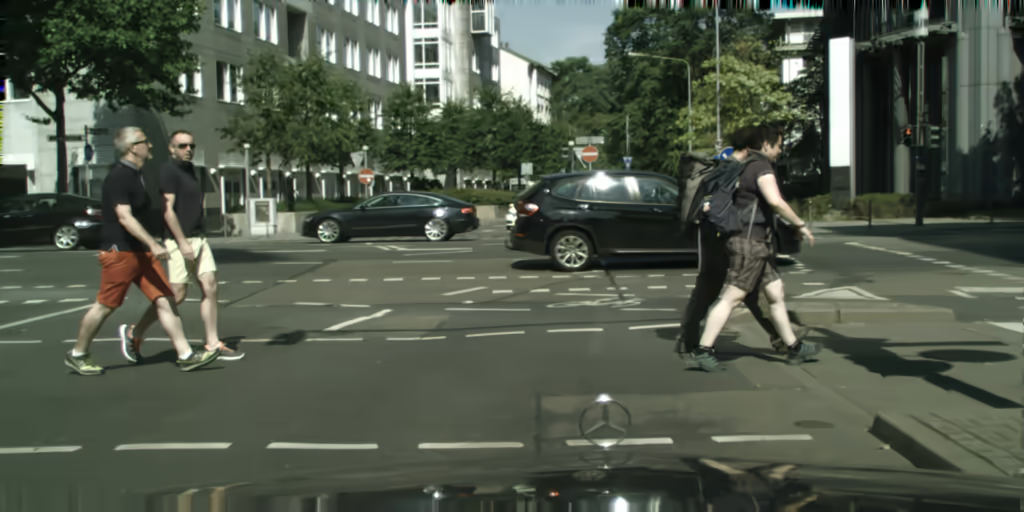}}
			\hspace{-.20in}
			\caption{Generating different instances for one image using our CCT model at SNR=$10$dB. In (b)-(e), the subtracted instance (e.g. ``--\emph{car}'') indicates the corresponding instance is generated (not transmitted), and we additionally generate the instances \emph{vegetation}, \emph{sky}, \emph{sidewalk}, \emph{wall}. Red number and blue number indicate the percentage of bandwidth cost increase and saving compared to the CCT (all transmitted) scheme.}\label{fig13}
		\end{center}
		\vspace{-1em}
	\end{figure*}
	
	\subsection{Results of CCT}\label{Results_Analysis_CCT}
	We first evaluate the interactive user-server process based on CCT (described in \ref{Interactive_and_Scalable}), where we report the FID and PSNR performance on Cityscapes dataset over the AWGN channel at SNR = $10$dB in Fig. \ref{fig11} (a) and (b). During this process, the users can interact multiple times until they are satisfied with the perceptual quality of received images. For simplicity, we pick four typical instances (\emph{car}, \emph{road}, \emph{people}, \emph{building}) as optional prompts for the user, and we restrict a prompt to include only one instance. Thus, one user can have up to 4 interactions, and we evaluate the reconstructed images in the transmissions for all the 15 combinations from \{\emph{car}, \emph{road}, \emph{people}, \emph{building}\}. Besides, we evaluate the images recovered from only instance label maps (masking all the instances) and all instances (not masking). In Fig. \ref{fig11} (a) and (b), each colorful data point denotes one combination, and the dashed line connecting two data points denotes the added instance. From the results, as expected, the PSNR performance of CCT is inferior compared to ``VTM + LDPC''. However, the FID performance of all the CCT data points significantly outperforms ``VTM + LDPC'' by a large margin. In addition, we find that our FID performance is almost saturated when the number of preserved instances increases to 4.
	
	Figure \ref{fig11} (c) presents the mean IoU in relation to CBR on the Cityscapes validation set for CCT models over the AWGN channel at SNR = $10$dB, along with the results obtained for the DPCT model, ``VTM + LDPC'' and NTSCC+. The quantitative evaluation of the semantic preservation ability demonstrates that CCT models uphold content consistency marginally better than DPCT. This suggests that the CCT effectively generates lost contents while seamlessly integrating them with preserved regions. Attributed to blocky and blurry artifacts, the ``VTM + LDPC'' and NTSCC+ are significantly inferior to those obtained by our CCT. Comparing the CCT themselves, we find that the mean IoU decreases obviously when the preserved instances change from \{\emph{road}, \emph{car}, \emph{people}\} to \{\emph{building}, \emph{car}, \emph{people}\}, as \emph{building} has more sophisticated structures.
	
	In Fig. \ref{fig12}, the user study results of our CCT models at two CBRs on Cityscapes validation set at SNR = $10$dB are presented. Here, we only transmit \emph{car}, \emph{people}, and \emph{building}. In particular, we require that raters allow a certain degree of distortion for the synthesized regions, as they are unimportant to the user. From the results, our CCT models with CBR = $0.009$ and CBR = $0.0192$ are preferred to ``VTM + LDPC'', despite the images produced by ``VTM + LDPC'' consuming 142\% and 113\% more CBRs compared to those produced by our models.
	
	In Fig. \ref{fig13}, we display some example images produced by the CCT model at SNR = $10$dB, where different instances are generated. Our CCT models manage to merge preserved and synthesized image content seamlessly. Furthermore, our CCT reduces CBR by up to 46\% compared to the same network without generation, with no significant impact on visual quality. When generating these contents with repetitive structures (e.g. sky, vegetation, and roads), the perceptual quality is essentially unimpaired. We can see that Fig. \ref{fig13}(d) is visually more pleasing than NTSCC+ and ``VTM + LDPC'', while they use 110\% and 118\% more CBRs than Fig. \ref{fig13}(d).

	\section{Conclusion}\label{Conclusion}
	
	This paper has introduced the first RDP jointly optimized generative JSCC framework designed to enhance perceptual quality in human communications. By integrating a flexible plug-in conditional GAN, the RDP-JSCC framework provides high-fidelity image reconstructions while supporting controllable generative transmission. Based on this framework, we propose two RDP implementations: a realism map-assisted DPCT model that considers personalized DP preferences for bandwidth-affordable scenarios and an interest-oriented label map-assisted CCT model for bandwidth-scarce scenarios. Comprehensive experiments demonstrate the superiority of our RDP-JSCC framework: our DPCT model significantly outperformed existing transmission methods in terms of perceptual performance, and its distortion performance also approached that of ``VTM + LDPC" scheme in some cases. Additionally, our CCT model enabled a novel interactive and scalable image transmission system, maintaining remarkable perceptual performance while saving transmission bandwidth considerably.

	\ifCLASSOPTIONcaptionsoff
	\newpage
	\fi
	
	\bibliographystyle{IEEEtran}
	
	\bibliography{Ref}
	
\end{document}